\documentclass[preprintnumbers,twocolumn,amsmath,amssymb,floatfix,15pt,prd,onecolumn,
superscriptaddress,nofootinbib]{revtex4}
\usepackage{latexsym}
\usepackage{epsfig}
\usepackage{amssymb}
 \usepackage{amsmath}

\begin{document}

\title{ Gravitational  lensing due to charged galactic wormhole  }
\author{Md Khalid Hossain}%
\email{mdkhalidhossain600@gmail.com }
\affiliation{%
Department of Mathematics, Jadavpur University, Kolkata 700032, West Bengal, India}

\author{Farook Rahaman}
\email{rahaman@associates.iucaa.in}
\affiliation{Department of Mathematics, Jadavpur University, Kolkata 700032, West Bengal, India}

\date{\today}

\begin{abstract}
 We propose  the back reaction to the    charged galactic wormhole spacetime based on Yoshiaki Sofue's exponential dark matter density profile to find exact solutions. The charges act as an additional component to the static wormhole, which is primarily formed by the galactic dark matter density.
 Unlike traditional mass-based models, this solution incorporates charge effects within a realistic dark matter distribution, revealing unique interactions between dark matter, electromagnetic fields, and spacetime curvature.
     This study confirms the criteria for wormhole formation, designating it the "Charged Galactic Wormhole," and offers a new framework for investigating galactic structures, with potential observational signatures that deepen our understanding of dark matter and spacetime.  Later, the proper radial distance and the embedding surface were also analyzed. Furthermore, the deflection of light around a charged galactic wormhole was investigated, along with a comprehensive review of the resulting image. The deflection of massive objects (charge less) near charged galactic wormholes is studied using the Gauss-Bonnet and Rindler-Ishak methods, with a detailed comparison of the results from both approaches. Additionally, in both the Rindler-Ishak (RI) and Gauss-Bonnet (GB) methods, when $v$ tends to $1$  i.e. when particle's velocity comparable to the speed of light , the results from these approaches converge, producing the same outcome as strong gravitational lensing.

\end{abstract}

\pacs{04.40.Nr, 04.20.Jb, 04.20.Dw}
\maketitle
  \textbf{Keywords : } Strong lensing ;  Jacobi metric ; Charged wormhole; Deflection angle

\section{Introduction:}

Wormholes remain a promising field in observational astrophysics. In order to obtain data through observational detection, one must engage with the wormhole's geometrical structure. Therefore, in an effort to gain insight into its geometrical properties, scientists are attempting to visualize wormholes.
Gravitational lensing is among the initial experimental confirmations of the general theory of relativity \cite{r1}.First, it was recognized in the manner in which the sun deflects light, then in the manner in which foreground galaxies lens quasars, then in the way that massive arcs develop in galaxy clusters, and ultimately with the technique of galactic microlensing. It is now a common occurrence in the field of observatory astronomy \cite{r2}.
Fuller and Wheeler first proposed the idea of a non-simply interconnected space-time along with the concept of wormholes \cite{r3}. Just the same, as Einstein and Rosen had already investigated the Schwarzschild and Reissner-Nordstrom solutions' non-singular coordinate patches \cite{r4}.
The concept of traversable wormholes containing exotic stuff was later refined by Moris and Thorne \cite{r5}. Other researchers carried out more modifications \cite{r6,r7,r8}.
Lensing in the strong gravitational field regime is a relatively recent area of research that has garnered significant interest in recent times and also weak lensing from celestial objects has been thoroughly explored over the past century \cite{r9,r9a,r10,r11,r12,r13,r14,r15,r16}. While exotic matter is necessary for wormhole models to exist in General Relativity, it is now established that wormhole models can also exist in modified gravity theories with regular matter \cite{r17}-\cite{r46}.
Expanding these study avenues is highly desirable, especially in view of the initial findings from the Event Horizon Telescope \cite{r47,r48}.
Indeed, a wealth of knowledge on the BH signatures through lensing process has been revealed by all these experiments.Still, there's another intriguing idea that hasn't gotten enough attention up until this point: Just as fascinating as black holes (BHs), it involves lensing caused by stellar-scale traversable wormholes (WH). WHs, which have throats connecting two asymptotically flat regions of spacetime, have been the subject of much research in the literature.
   Recently, it is proposed that wormholes could be existed ingalactic regions \cite{r48a,r48b}.  Yoshiaki Sofue \cite{r48c}  introduced a novel dark matter density profile for certain galaxies, referred to as the exponential density profile. This model describes the distribution of dark matter within galaxies and, in some cases, aligns more closely with observed galactic rotation curves. The exponential profile resembles the density distributions of stellar disks in spiral galaxies, hinting at a possible connection between the distributions of visible and dark matter. It is particularly useful for analyzing the dark matter concentration in the inner regions of galaxies, including the galactic center, where conventional models often face challenges.
We propose examining the backreaction on the charged galactic wormhole spacetime, utilizing Yoshiaki Sofue's exponential dark matter density profile to obtain exact solutions. Our proposed charged galactic wormhole metric challenges conventional thinking by deviating from traditional mass dominated models. At first glance, one might assume that a wormhole metric without a traditional mass term would struggle to explain familiar gravitational phenomena. However, this deviation actually opens a new frontier for studying non-traditional effects. By incorporating Yoshiaki Sofue's exponential dark matter density profile, the model may predict some important phenomena.
 Among its many intriguing consequences is light propagation, which is particularly noticeable in the Morris-Thorne-Yurtsever (MTY) WHs pacetime \cite{r50}.
Recent research on WH lensing was conducted by Safonova et al. \cite{r51}, following the recent work of Cramer et al. \cite{r52} on the issue of lensing by negative mass WHs. Morris-Thorne type WHs typically behave as convergent lenses, according to Tejeiro and Larranaga's most recent research \cite{r53}.
In the fields of astrophysics and cosmology, gravitational lensing serves as an invaluable tool for a number of tasks, including measuring the Hubble constant, mapping the distribution of dark matter, confirming the existence of extrasolar planets, calculating the cosmological constant, and understanding mass distribution in the vast structure of the universe \cite{r55}.

The structure of this paper is organized as follows: After solving the Einstein-Maxwell equations in Section \ref{1}, we obtain a charged galactic wormhole. In Section \ref{S}, we have discussed about feasibility of the wormhole. We looked into the wormhole's embedding surface and proper radial distance in Section \ref{2}. We have now looked into the photons' deflection nearby the wormhole in Section \ref{3}. The entire image analysis and einstein ring were covered in Section \ref{4}. Lastly, we have examined the deflection of massive object around the wormhole using the Rindler-Ishak and Gauss-Bonnet methods in sections \ref{5} and \ref{6}. We have drawn a general conclusion on the study in Section \ref{7} later on.

\section{Charged galactic wormhole} \label{1}

Assume that the Einstein equation applies to the simplest form of conventional wormhole
\begin{equation}
    G^{(0)}_{\mu\nu}=8\pi T^{(0)}_{\mu\nu}
\end{equation}

The term $G^{(0)}_{\mu\nu}$ represents the geometry of the wormhole, while the term $T^{(0)}_{\mu\nu}$ corresponds to exotic matter that violates established energy conditions. The wormhole cannot be established without this exotic matter.

 {Following Kim et al \cite{rnew1}, one can introduce  back reaction by introducing the additional matter $T^{(1)}_{\mu\nu}$ on the right side and encompassing the related backlash $G^{(1)}_{\mu\nu}$ on the left side. Now, the equation takes the form:}

\begin{equation}
    G^{(0)}_{\mu\nu}+G^{(1)}_{\mu\nu}=8\pi \left(T^{(0)}_{\mu\nu}+T^{(1)}_{\mu\nu}\right).
\end{equation}

The total matter on the right side naturally obeys conservation law. Although, this does not ensure that the matter as a whole remains exotic. When the backlash on the wormhole's geometry, $G^{(1)}_{\mu\nu}$, becomes sufficiently large to take over the system, the extra material might make it impossible for the wormhole to endure.

 {A wormhole having an electric charge could possibly be defined as follows:
\begin{equation}\label{e1}
    ds^{2}=-\left(1+\frac{Q^2}{r^2}\right)dt^2 + \left(1-\frac{b(r)}{r}+\frac{Q^2}{r^2}\right)^{-1}dr^2+r^2(d\theta^2+\sin^2\theta\ d\phi^2).
\end{equation}
This spacetime represents a combination of the Morris-Thorne (MT) spherically symmetric static wormhole and the Reissner-Nordström (RN) spacetime. When
$
Q=0$, the spacetime corresponds to the MT wormhole, and when
$
b=0$, it reduces to the RN black hole with zero mass.}

The Einstein-Maxwell equations take the form:

\begin{equation}\label{k1}
    \frac{b'}{r^2}+\frac{Q^2}{r^4}=8\pi\left(\rho^{(0)}+\rho^{(1)}\right),
\end{equation}

\begin{equation}\label{k2}
    \frac{b}{r^3}-\frac{Q^2}{r^4}-2\left( 1-\frac{b}{r}+\frac{Q^2}{r^2}\right)\left(-\frac{Q^2}{r^2(r^2+Q^2)}\right)=8\pi\left(\tau^{(0)}+\tau^{(1)}\right),
\end{equation}

\begin{multline}\label{k3}
  \left(1-\frac{b}{r}+\frac{Q^2}{r^2}\right)\Biggr[\frac{Q^2(3r^2+Q^2)}{r^2(r^2+Q^2)}+\left(\frac{b'r-b+\frac{2Q^2}{r^2}}{2(r^2-br+Q^2)}\right) \left(\frac{Q^2}{r(r^2+Q^2)^2}\right)+\left(-\frac{Q^2}{r(r^2+Q^2)^2}\right)^2-\left(-\frac{Q^2}{r(r^2+Q^2)^2}\right)\\
  -\left(\frac{b'r-b+\frac{2Q^2}{r^2}}{2(r^2-br+Q^2)}\right)\Biggr] =8\pi\left(P^{(0)}+P^{(1)}\right).
\end{multline}

Here, the prime symbol indicates differentiating with the variable $r$. Considering a geometry that is spherically symmetric, the elements of $T^{(0)}_{\mu\nu}$ can be determined in orthonormal systems.
\begin{equation}
T^{(0)}_{tt}=\rho^{(0)}(r),
T^{(0)}_{rr}=-\tau^{(0)}(r),
T^{(0)}_{\theta\theta}=P^{(0)}(r).
\end{equation}

Here, $\rho^{(0)}(r) \rightarrow$  energy density ; $\tau^{(0)}(r) \rightarrow $ radial tension per unit area; $P^{(0)}(r) \rightarrow$ lateral pressure. For a viewer at a particular constant $r$, $\theta$, and $\phi$.

The matter terms due to the present electric field only are

\begin{equation}
  \rho^{(1)}=\tau^{(1)}=P^{(1)}=\frac{Q^2}{8\pi r^4} .
\end{equation}

Recently, Yoshiaki Sofue  \cite{r48c} introduced a new dark matter density profile for spiral galaxies, referred to as the exponential density profile, which is expressed as:

\begin{equation}
    \rho^{w}=\rho_{s}e^{-\frac{r}{r_s}}.
\end{equation}
Here, $r_s$ denotes scale radius, and $\rho_s$ represents central density.
Now we assume that $\rho^{(0)}=\rho^{w}$.
So we have
\begin{equation}
 \rho^{(0)}=\rho^{w}=\rho_{s}e^{-\frac{r}{r_s}}.
\end{equation}

By substituting the values of $\rho^{(0)}$ and $\rho^{(1)}$ into Eq.(\ref{k1}) and solving it, we obtain
\begin{equation}
b(r)=-8r_s\left(r^2+2rr_s+2r^2_{s}\right)e^{-\frac{r}{r_s}} \pi \rho_{s}+C_1 ,
\end{equation}
where $C_1$ is constant.

By substituting the values of $\tau^{(1)}$ and $b(r)$ into Eq.(\ref{k2}) and solving it, we obtain $\tau^{(0)}$.\\

Similarly, by substituting the values of $P^{(1)}$ and $b(r)$ into Eq.(\ref{k3}) and solving it, we obtain $P^{(0)}$.

Hence the metric (\ref{e1})     satisfies  the  Einstein's  equation  self-consistently. In a static, spherically symmetric spacetime, the
$
g_{
tt}
$
  component represents the gravitational potential, influencing both particle motion and gravitational redshift. The redshift arises due to this potential, as described by
$
g_{
tt}
$
 . Even in the absence of a conventional mass term in a wormhole metric, the
$
g_{
tt}
$
  component can still generate a gravitational potential capable of causing redshift effects.

\section{Feasibility of wormhole }\label{S}
\begin{itemize}

\item Our shape function is:
\begin{equation}\label{w1}
    b_{eff}(r)=b(r)-\frac{Q^2}{r}.
\end{equation}

According to \cite{r58a,r58b}, the shape function (\ref{w1}) must meet the following requirements in order to preserve the wormhole structure.

The shape function derived here is positive and monotonically growing in nature following the wormhole throat, as can be shown in Fig. (\ref{a}) (left side) .

To further illustrate this, $\frac{b_{eff}(r)}{r}< 1$ ( left side of Fig. (\ref{b})) and $\frac{db_{eff}(r)}{dr}< 1$ ( right side of Fig. (\ref{b})) are presented for all $r>t_{th}$ (wormhole throat ).

Therefore, all the requirements including  flaring-outward constraint
are satisfied  for building the wormhole structure.
\begin{figure}[h]

\includegraphics[width=8.5cm]{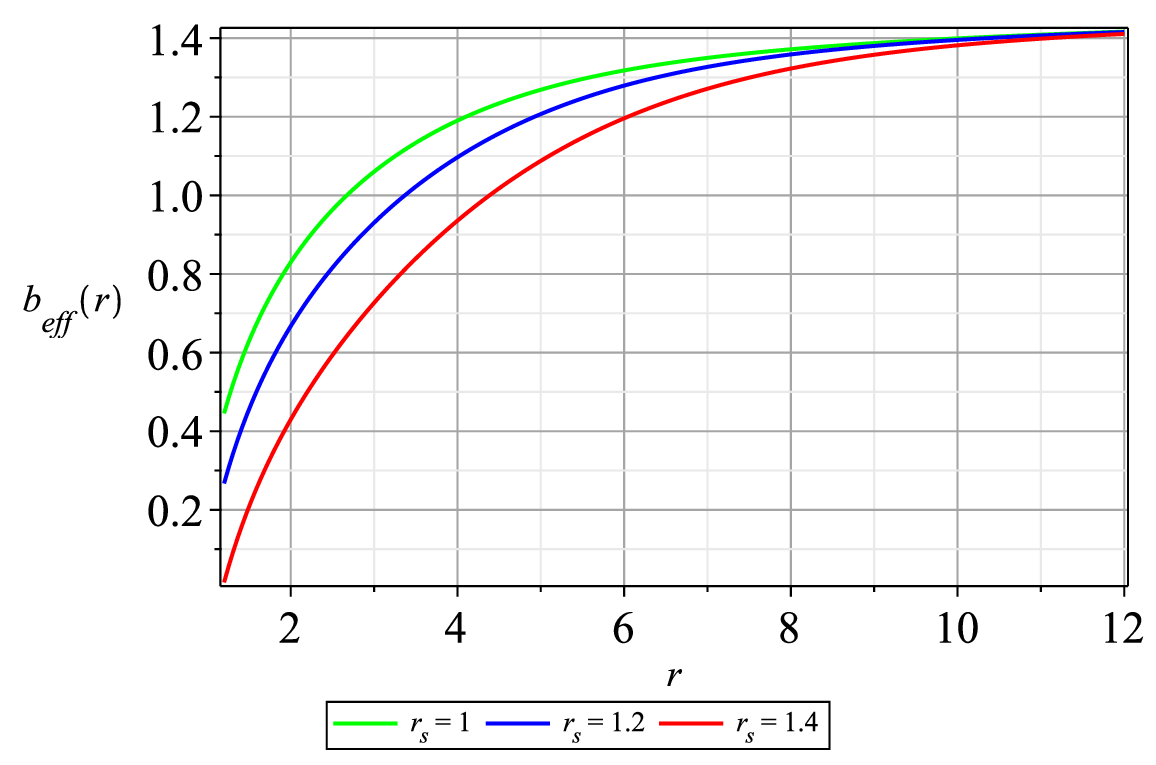}
\includegraphics[width=8.5cm]{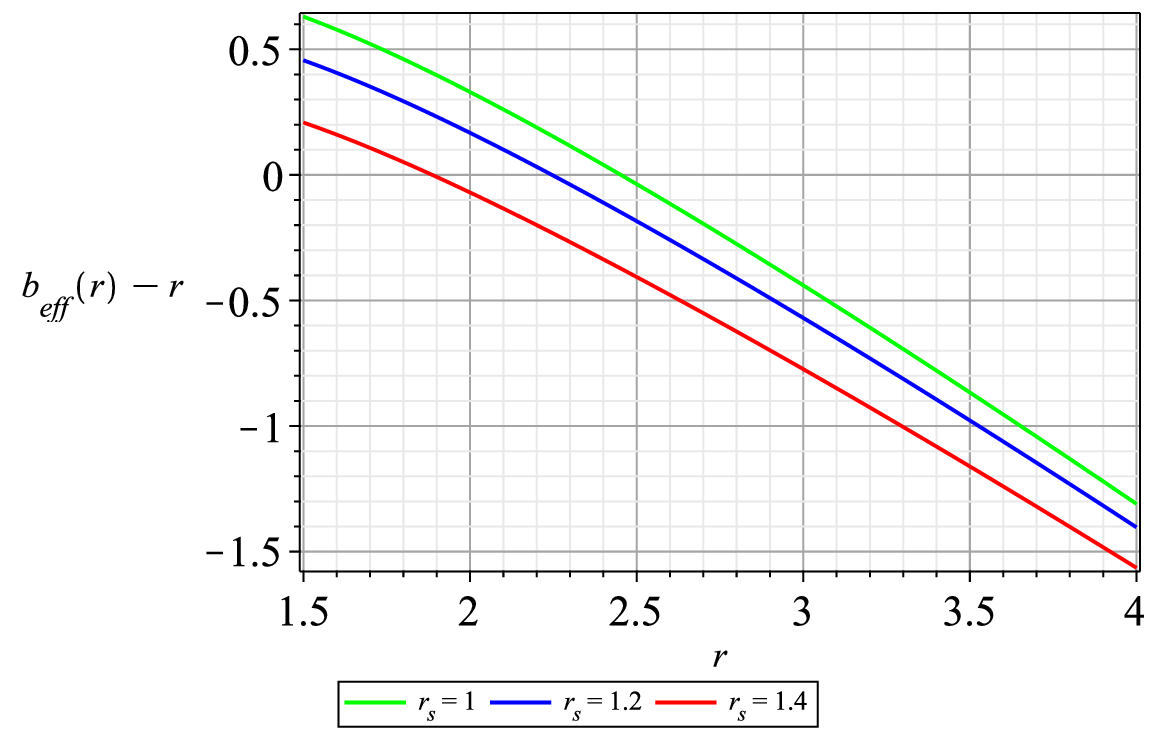}
\caption{The above diagram is   the graphical representations of   $b_{eff}(r)$  for the various values of $r_s$ (left) and radius of the throat where $b_{eff}(r) -r $  cuts r axis (right). } \label{a}
\end{figure}

\begin{figure}[h]
\includegraphics[width=8.5cm]{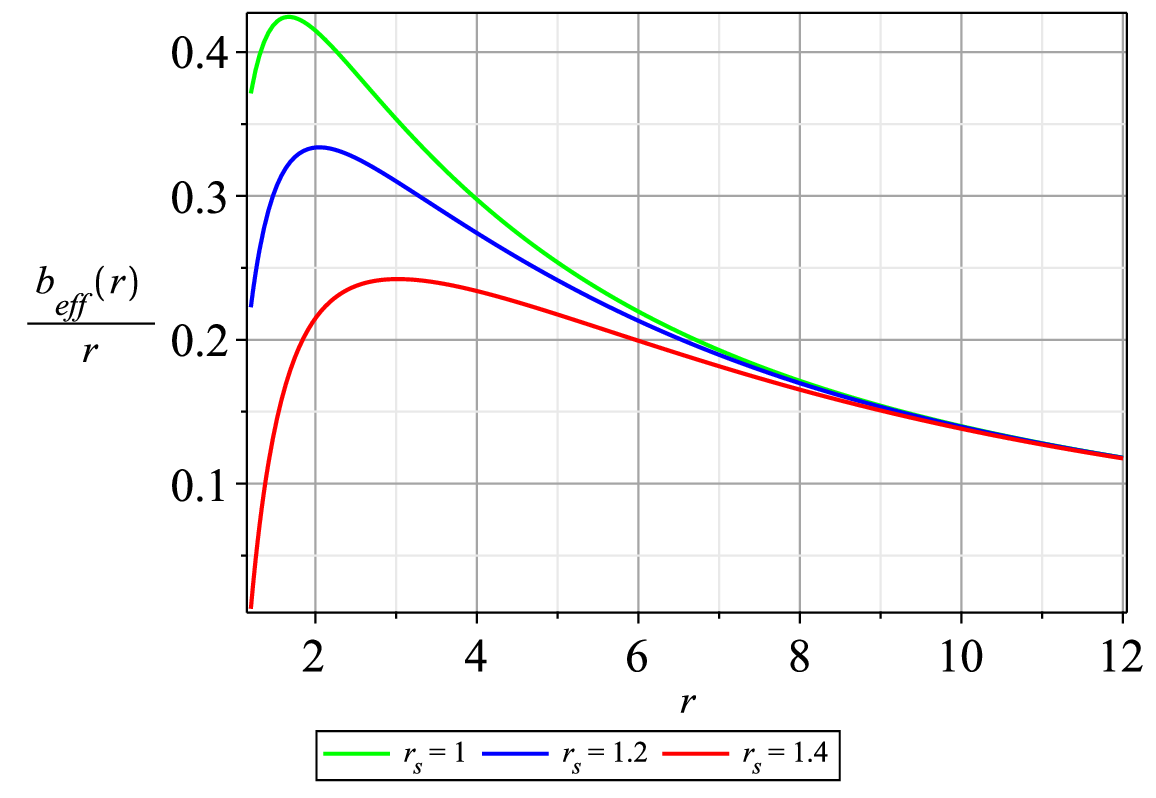}
\includegraphics[width=8.5cm]{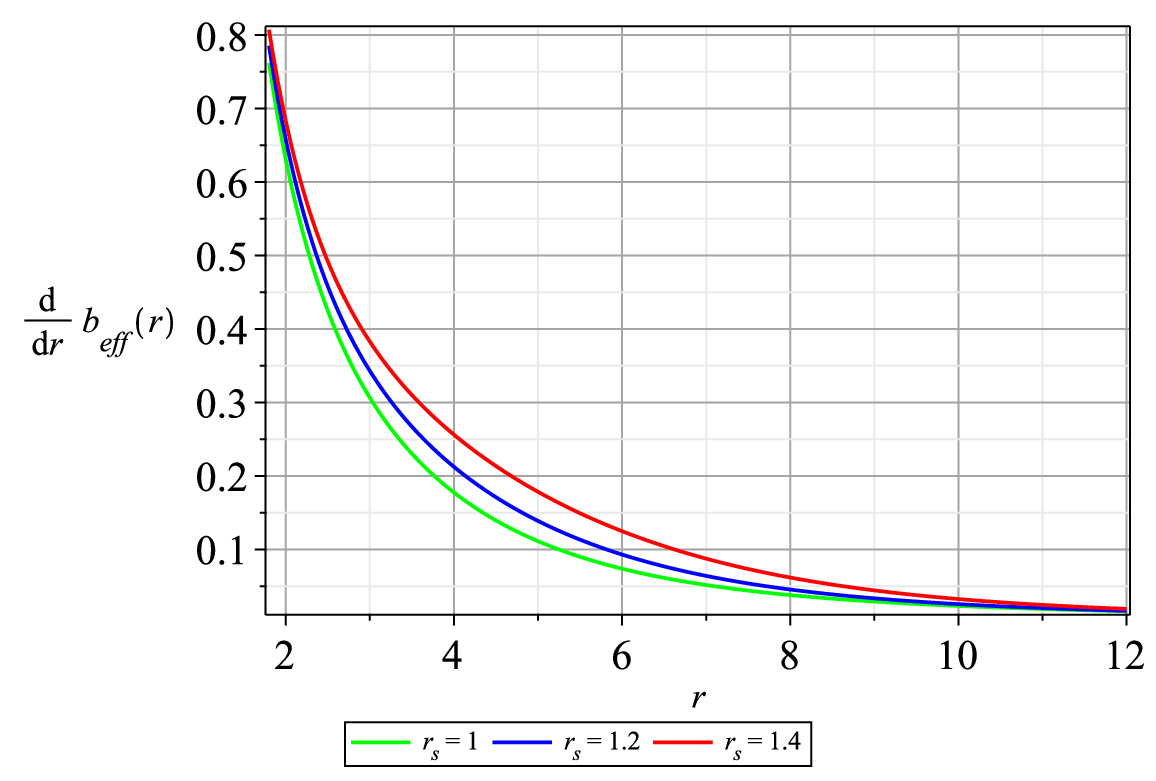}
\caption{ The above diagrams are the graphical representations of $\frac{b_{eff}(r)}{r}$ and $\frac{db_{eff}(r)}{dr}$ in relation to $r$ in left and right part, respectively for the various values of $r_s$.  } \label{b}
\end{figure}

\item Based on the findings in \cite{r58a, r58b}, we also analyze the matter composition of the wormhole to assess whether it violates the null energy condition (NEC), a crucial factor in sustaining the wormhole's stability. According to the General Theory of Relativity (GTR), the NEC is defined as follows:

 $ \rho(r) + P_r(r) \geq 0$  or $ \rho(r) + P_t(r) \geq 0$

where $\rho(r)$ is the energy density, $P_r(r)$ is the radial pressure, and $P_t(r)$ is the tangential pressure. The NEC is violated by the wormhole's substance content if\\
 $ \rho(r) + P_r(r) < 0,  \quad \text{for} \quad r \geq r_{\text{th}}.$\\
or
$ \rho(r) + P_t(r) < 0,  \quad \text{for} \quad r \geq r_{\text{th}}.$
\begin{figure}[h]
\includegraphics[width=8.5cm]{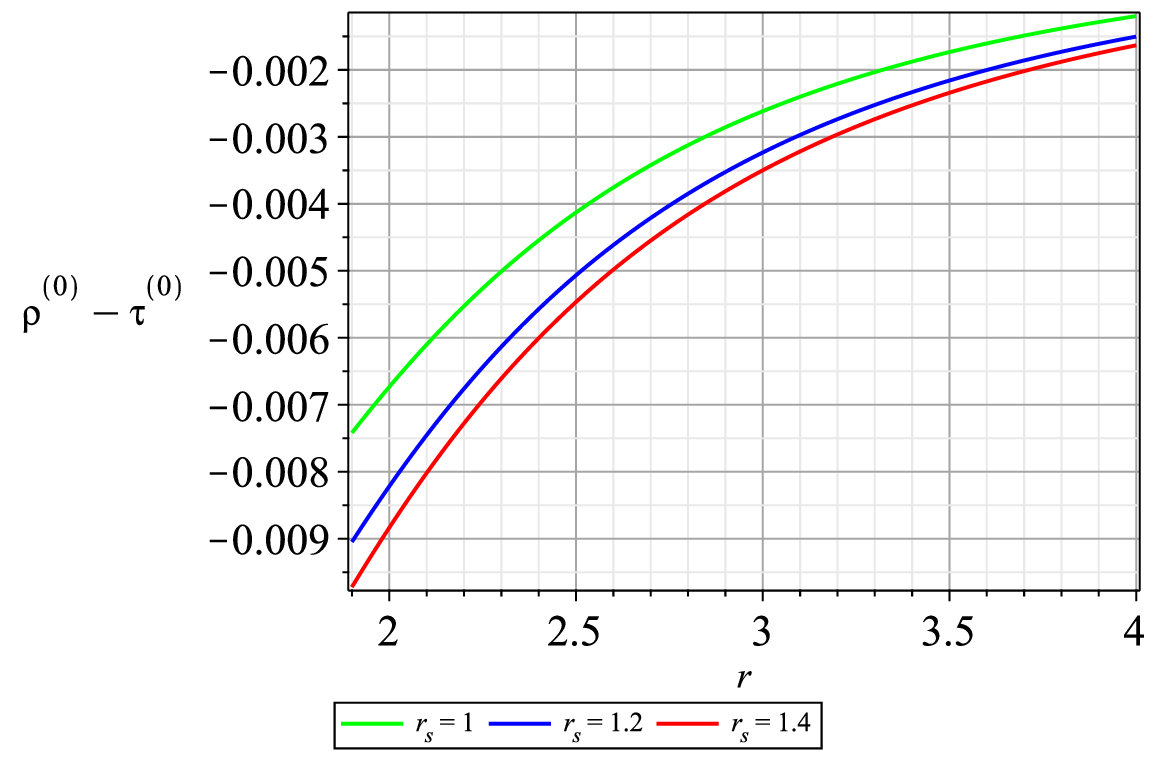}
\includegraphics[width=8.5cm]{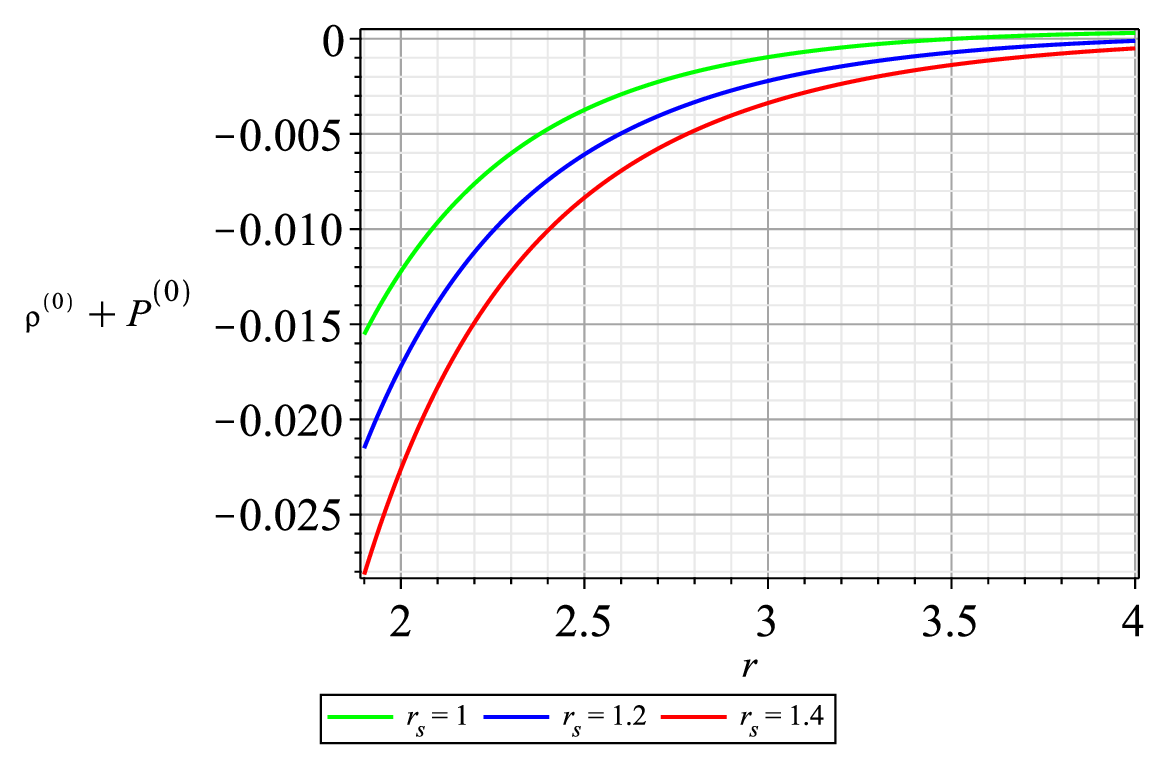}
\caption{ The above diagrams are the graphical representations of $\rho^{(0)}-\tau^{(0)}$ (i.e., energy density + radial pressure) and
 $\rho^{(0)}+P^{(0)}$ (i.e., energy density + tangential pressure) in relation to $r$ in the left and right part, respectively for the various values of $r_s$. } \label{c}
\end{figure}

In the left side Fig. (\ref{c}) as well as right side , we can see that $\rho^{0}-\tau^{0}<0,  \quad \text{for} \quad r \geq r_{\text{th}} $
 and   $\rho^{0}+P^{0}<0,  \quad \text{for} \quad r \geq r_{\text{th}}.$\\
Hence the NEC is violated to have a wormhole configuration.
Thus, all viability criteria   are satisfied for the wormhole structure.

\item Now, we will check whether the stability of our wormhole by using causality conditions.
  We use the squared adiabatic sound speed method provided by \cite{r58b} to verify the wormhole's stability.
\begin{equation}
    v^2_{s}=\frac{dP}{dr}\left(\frac{d\rho}{dr}\right)^{-1},
\end{equation}
where, $\rho$ is energy density, $P$ represents both  radial pressure and  tangential pressure.

In Fig. (\ref{ww}), one clearly sees that   $0 \leq v^2_{s} < 1$ , where $v^2_{s}= v^2_{sr} ~or~ v^2_{st}$. Thus radial and transverse sound speed  is smaller than the speed of light. Thus causality constraint is satisfied which  imply  that our wormhole is stable.

We have assumed the  particular values  as $Q=1$, $r=1.5$, $\rho_{s}=0.005$ and $C_{1}=0$. In order to consider the causality condition, we need to impose the following constraint on $r_{s}$:
$0.1\leq r_{s}\leq 1.9$.

\begin{figure}[h]
\includegraphics[width=8.5cm]{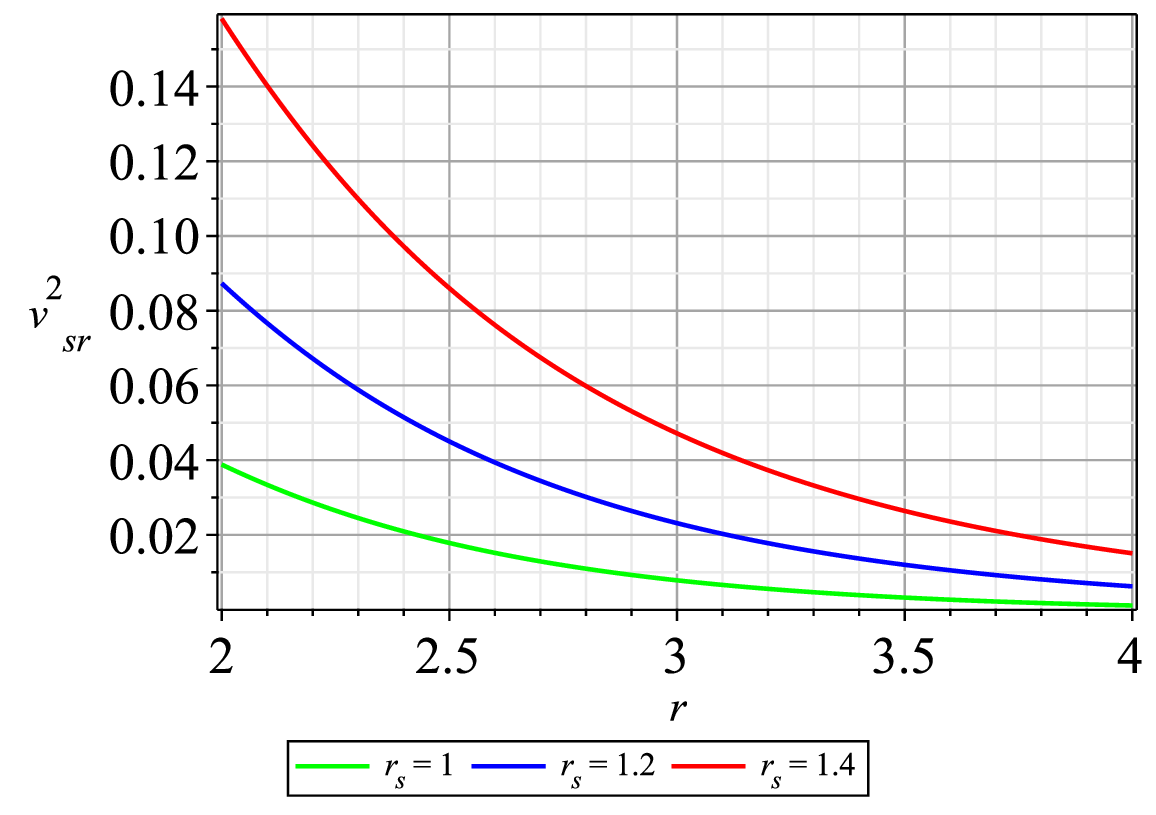}
\includegraphics[width=8.5cm]{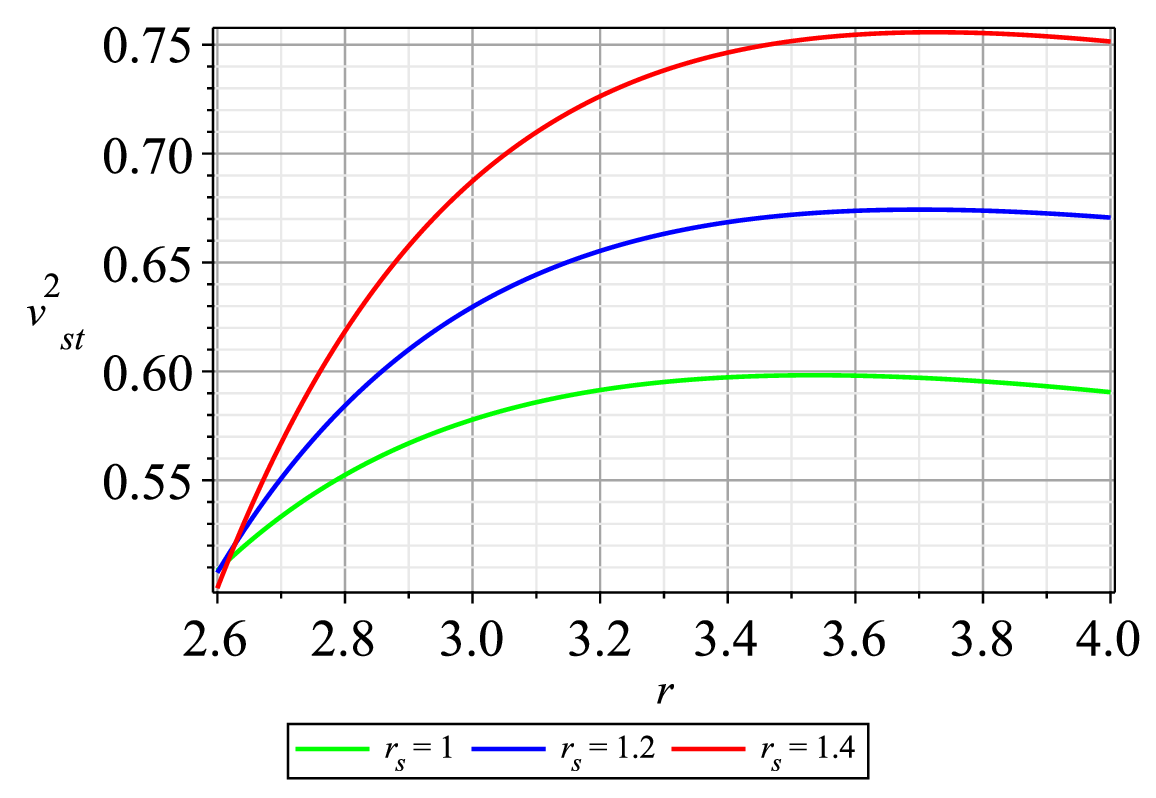}
\caption{The above diagrams are the graphical representations of $v^2_{sr}$ and $v^2_{st}$ in relation to $r$ in the left part and right part, respectively the various values of $r_s$. } \label{ww}
\end{figure}

\end{itemize}

\clearpage

\section{Study of Surface Embedding and Proper Radial Distance} \label{2}

The embedding surface and the wormhole's proper radial distance are two more crucial aspects in the physical principles of wormhole geometry. The wormhole's embedding surface is represented via $z(r)$ function, that corresponds with the subsequent formula \cite{r58}
\begin{equation}\label{k10}
    \frac{dz(r)}{dr}=\pm\left(\frac{r}{b_{eff}(r)}-1\right)^{-\frac{1}{2}},
\end{equation}
where, $b_{eff}(r) = b(r) - \frac{Q^2}{r} $.\\

The divergence of $\frac{dz(r)}{dr}$ at the wormhole throat in Eq.(\ref{k10}) indicates the embedding surface gets vertical at the wormhole throat. The embedding surface $z(r)$ integral expression is obtained by applying Eq.(\ref{k10}) to the data.

\begin{equation}\label{k11}
    z(r)= \pm \int_{r^+_{th}}^{r}\left(\frac{r}{b_{eff}(r)}-1\right)^{-\frac{1}{2}} \,dr .
\end{equation}

The proper radial distance is represented by $l(r)$, which is stated as
\begin{equation}\label{k12}
    l(r)= \pm \int_{r^+_{th}}^{r}\left(1-\frac{b_{eff}(r)}{r}\right)^{-\frac{1}{2}} \,dr .
\end{equation}

\begin{figure}[h]
\includegraphics[width=8.5cm]{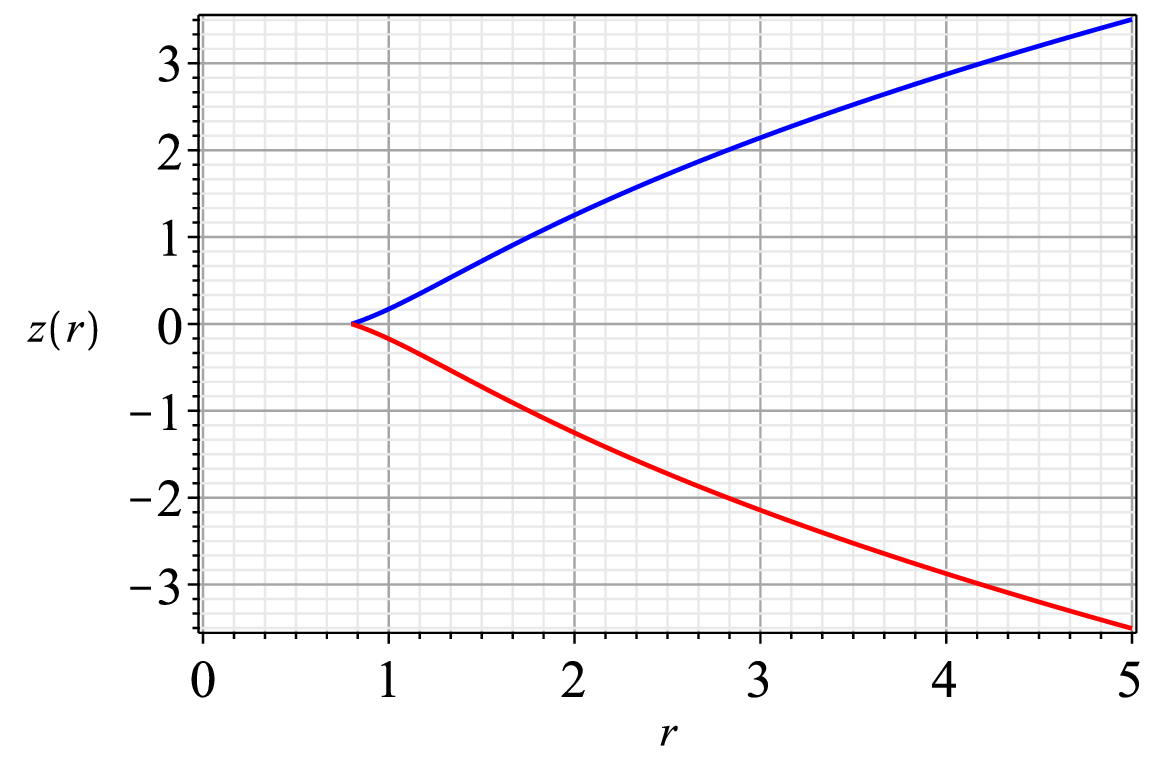}
\includegraphics[width=8.5cm]{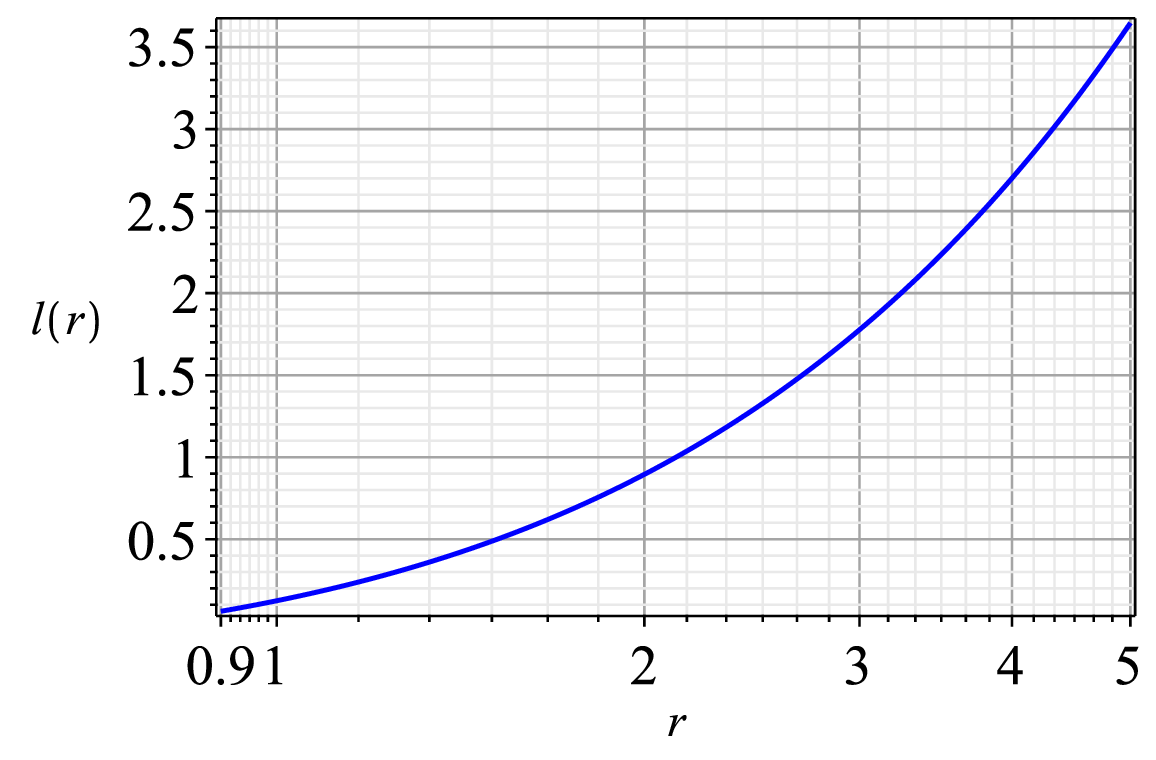}
\caption{The above diagrams are the graphical  representation of embedding surface $z(r)$ and radial distance $l(r)$ that correspond considering the criteria $Q=1$, $\rho_0=0.00003$ and $r_s=1$. .} \label{d}
\end{figure}

\begin{figure}[h]
\centering
\includegraphics[width=9cm]{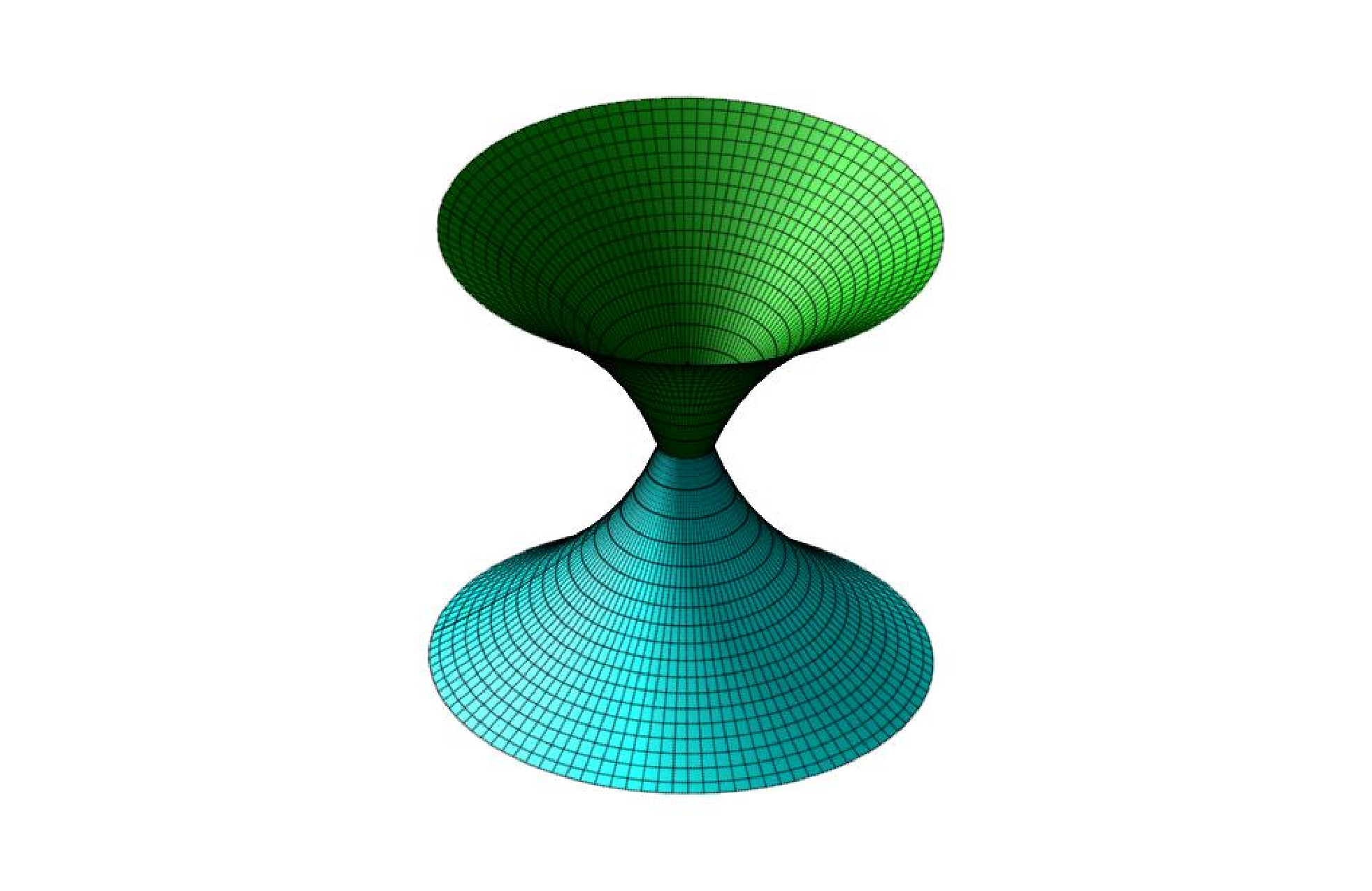}
\caption{ The above diagram is the full visualizing of the charged wormhole. } \label{e}
\end{figure}

\begin{table}[h]
\centering
\caption{Various values are provided to represent the accurate radial distance $l(r)$ and embedding surface $z(r)$ that correspond to the parameters $\rho_0=0.00003$ and $r_s=1$.\label{table}}
\begin{tabular}{|p{5cm} p{5cm} p{5cm}|}
\hline
$r(kpc)$ & $z(r)$ & $l(r)$ \\
\hline
0.85 & 0.0364145527791212 & 0.0296946790820933 \\
1.0 & 0.171580310322962 & 0.124595129892207 \\
1.5 & 0.723210321874316  & 0.487543872656603 \\
2.0 & 1.25348696054533 & 0.895225340712104 \\
2.5 & 1.72365797573364 & 1.32818941070741 \\
3.0 & 2.14349839150619 & 1.77670459088976 \\
3.5 & 2.52430870442539 & 2.23547788637023   \\
4.0 & 2.87449107199794 & 2.70140410175315 \\
4.5 & 3.20005433773007 & 3.17254229696396 \\
5.0 & 3.50536667780223 & 3.64761476648819 \\
\hline
\end{tabular}
\end{table}

where $r^+_{th}$ is the wormhole throat's closest right side distance.Solving the above integrals (\ref{k11}) and (\ref{k12}) using the analytical method is not feasible. Therefore, we eventually used the Maple software to perform numerical method for evaluating the integrals Eq.(\ref{k11}) and Eq.(\ref{k12}) in order to adjust the behaviors of $z(r)$ and $l(r)$.
It is important to observe that both numerical integrations are performed by adding a specific step length to change the upper limit and fixing a specific value of the lower limit, $r^+_{th}= 0.9$ kpc. A couple of data for
$z(r)$ and $l(r)$ are presented in Table \ref{table}, corresponding to the integrals' specified upper limit. Fig.(\ref{d}) illustrates the graphical representations of the radial proper distance $l(r)$ and the embedding surface $z(r)$, obtained using a very small integration step size. A complete illustration regarding the 4D wormhole, generated via the embedding surface's rotation $z(r)$ around the $Z-$axis, is shown in Fig.(\ref{e}) for our solution.

\clearpage

\section{STRONG GRAVITATIONAL LENSING
} \label{3}

 {Following the approach initiated by Virbhadra et al \cite{r9a}, we will discuss    gravitational lensing as a tool to find distinctive lensing features of wormholes.}

The metric coefficients in Eq.(\ref{e1}) are first written as follows to determine the deflection angle that light experiences as a result of the exponential wormhole's gravity.
\begin{equation}
    A(r):=\left(1+\frac{Q^2}{r^2}\right),
\end{equation}

\begin{equation}
   B(r):= \left(1-\frac{-8r_s\left(r^2+2rr_s+2r^2_{s}\right)e^{-\frac{r}{r_s}} \pi \rho_{s}+C_1}{r}+\frac{Q^2}{r^2}\right),
\end{equation}

\begin{equation}
    C(r):=r^2 .
\end{equation}
 {According to references \cite{r9,r9a}, in a static spherically symmetric spacetime, a photon sphere is defined as a timelike hypersurface at
$
r_{
ps}$
 , where the Einstein bending angle of a light ray becomes infinitely large as the closest approach distance
$
r_0
$
  approaches
$
r_{
ps}$
. Consequently, the photon sphere equation for a general static spherically symmetric metric can be derived from Eq. (\ref{e1}):}

\begin{equation}
\frac{C'(r)}{C(r)}- \frac{A'(r)}{A(r)}=\frac{2}{r}+\frac{2Q^2}{r^3\left(1+\frac{Q^2}{r^2}\right)} =0,
\end{equation}

which reduces to,
\begin{equation}
    r_{m}:=\sqrt{2}Q.
\end{equation}

 {The photon sphere is important because it shows us where light spirals around the  wormhole and the angle of deflection goes to infinity.
One can determine the relationship among the $u$ (impact parameter) and $r_0$ (closest approach) utilizing the angular momentum preservation law \cite{r11},}

\begin{equation}\label{k23}
    u=\sqrt{\frac{C(r_0)}{A(r_0)}.}
\end{equation}

From the geodesic equation, it is now possible to calculate the deflection angle as \cite{r11},

\begin{equation}
    \alpha(r_0)=I(r_0)-\pi ,
\end{equation}

whereas,
\begin{equation}
    I(r_0)=\int_{0}^{\infty}\left(\frac{2\sqrt{B}dr}{\sqrt{C}\sqrt{\frac{CA_0}{C_0A}-\frac{C}{C_0}}}\right).
\end{equation}

Here assume that $A(r_0)=A_0$,$B(r_0)=B_0$
and $C(r_0)=C_0$.

 {Now introduce parameter y and $y_0 = y(r_0)$, which are defined as follows, it will now be introduced as functions of our updated spatial variable $r$ as a function of two variables and an additional variable $z$ \cite{r11}}

\begin{equation}
    y:=A(r)=\left(1+\frac{Q^2}{r^2}\right),
\end{equation}

and
\begin{equation}
    z:=\frac{y-y_0}{1-y_0},
\end{equation}

which states,

\begin{equation}
    r=\frac{r_0}{\sqrt{1-z}}.
\end{equation}

Therefore, we are able to utilize the new variable $z$ in substitution of the old variable $r$ in the metric coefficients as

\begin{equation}
    A(z,r_0)=1+\frac{Q^2(1-z)}{r^2_{0}},
\end{equation}

\begin{equation}
     B(z,r_0)=\frac{1}{1+\frac{Q^2(1-z)}{r^2_{0}}+\frac{\left(8r_s\left(\frac{r^2_{0}}{1-z}+\frac{2r_{0}r_{s}}{\sqrt{1-z}}+2r^2_{s}\right)\pi \rho_{s}e^{-\frac{r_0}{r_{s}\sqrt{1-z}}}-C_1\right)\sqrt{1-z}}{r_0}},
\end{equation}

\begin{equation}
   C(z,r_0)=\frac{r^2_{0}}{1-z} .
\end{equation}

 {Using Virbhadra's   approach \cite{r9,r12}, we can calculate our wormhole's deflection angle   as follows}

\begin{equation} \label{k20}
    \alpha(r_0)=-a\log \left(\frac{r_0}{r_m}-1\right)+b_R+b_D-\pi+O(r_0-r_m).
\end{equation}

Through numerical integration, we have determined the coefficient $b_R$ from the regular part of the integral. Because of the length of expression,it has not been added here.

The lensing coefficients are as follows:

\begin{equation}
    a=\frac{(Q^2+r^2_{m})^{\frac{3}{2}}r_{m}\sqrt{2}}{Q^2\sqrt{(Q^2+r^2_{m}+40\pi\rho_{0}r^4_{m}e^{-1}-C_{1}r_{m})(2r^2_{m}-4Q^2)}},
\end{equation}

\begin{equation}
    b_D=0,
\end{equation}

and
\begin{equation}
    b=b_D+b_R-\pi .
\end{equation}

\begin{figure}[h]
\includegraphics[width=8.5cm]{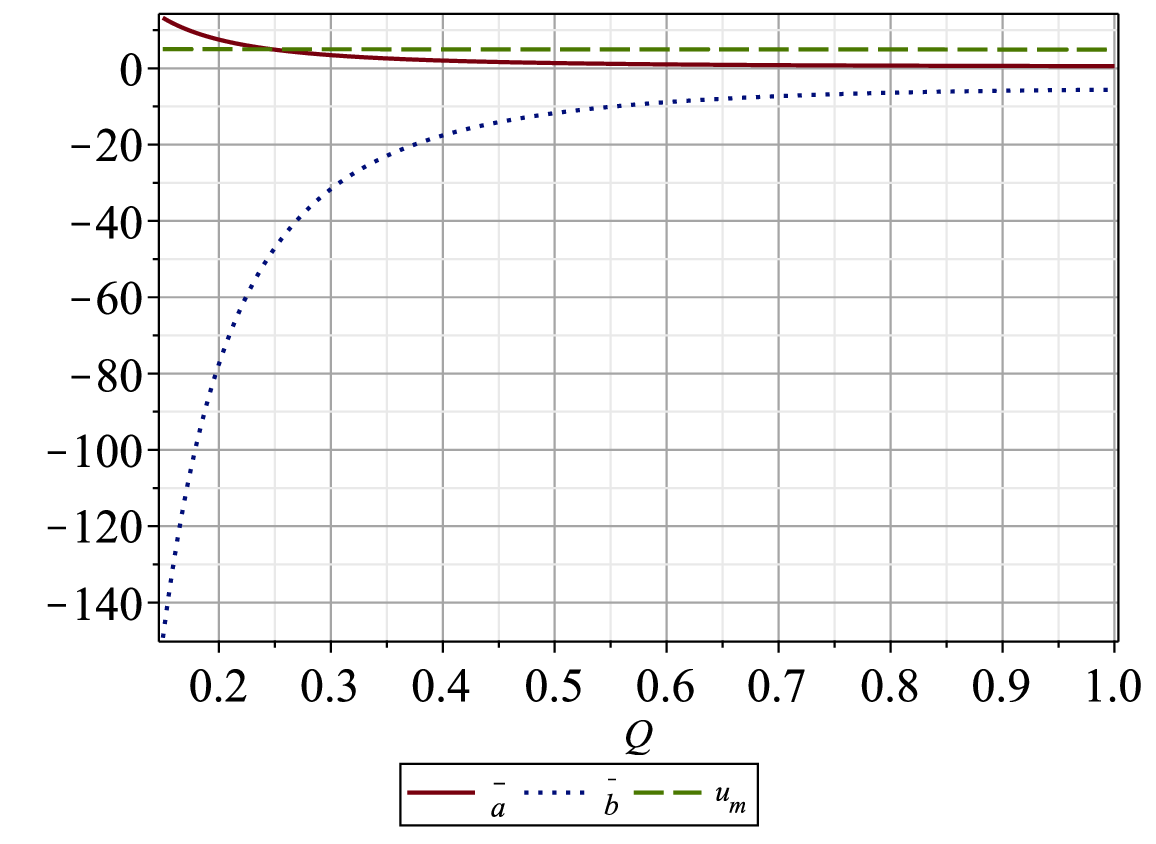}
\includegraphics[width=8.5cm]{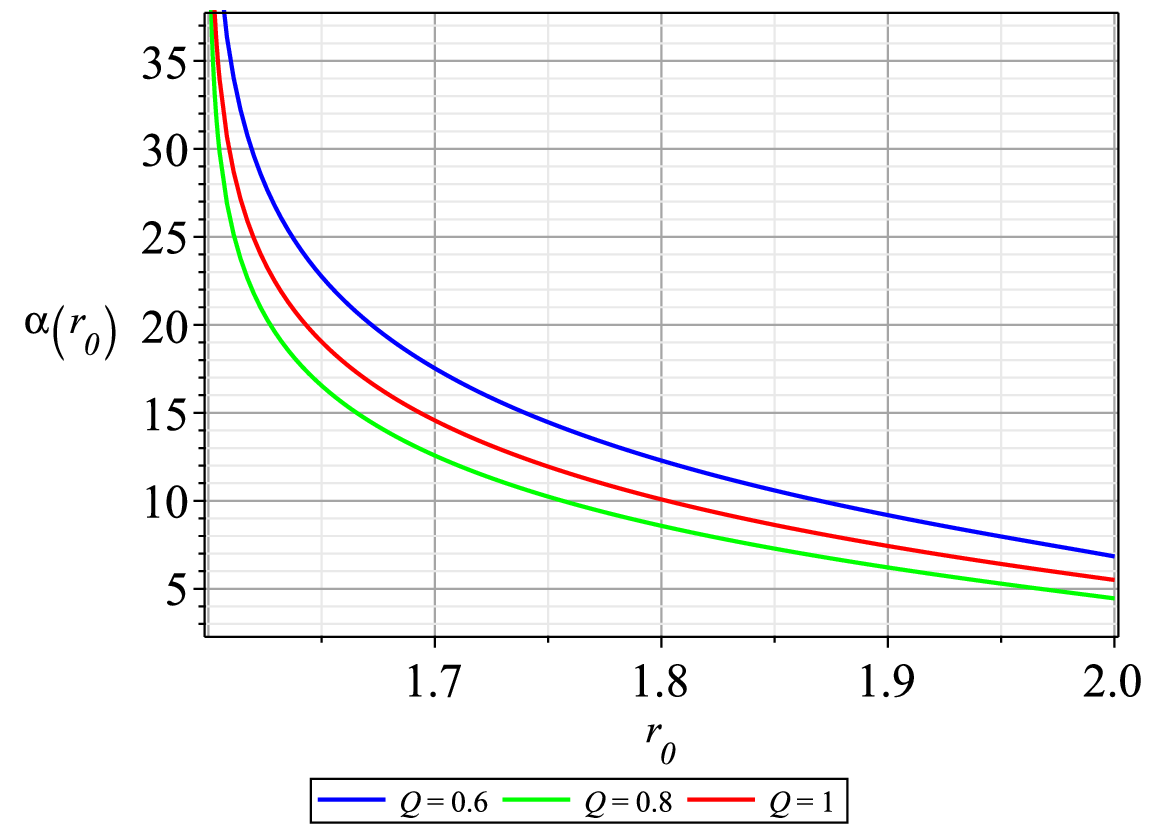}
\caption{ The above diagrams are the graphical  representation of the lensing coefficients $\bar{a}$, $\bar{b}$ and $\bar{u_m}$ as functions of parameter $Q$ (left part) and deflection angle $\alpha(r_0)$ in relation to the closest approach distance $(r_0)$ for various values of $Q$ (right part). } \label{f}
\end{figure}

In Fig.(\ref{f}) (left), we have shown coefficients $\bar{a}$, $\bar{b}$ and $\bar{u_m}$ as $Q$-dependent. Plotting the deflection angle $\alpha(r_0)$ against the closest approach $(r_0)$ to the wormhole, as provided by Eq.(\ref{k20}) for various parameter $Q$ values, is shown in the same figure's right panel.

Suppose the angle separations between the viewer and the lens are $D_{ol}$ and the image and the lens are $\theta=\frac{u}{D_{ol}}$
Eq.(\ref{k23}) gives

\begin{equation}
     u=\sqrt{\frac{C(r_m)}{y(r_m)}}=\frac{r^2_{m}}{\sqrt{Q^2+r^2_{m}}},
\end{equation}

Expanding Eq.(\ref{k23}) yields

\begin{equation}
    u-u_m=\left(\frac{r^2_{m}}{\sqrt{Q^2+r^2_{m}}}\right)(r_0-r_m)^2.
\end{equation}

We express,
\begin{equation}
    \bar{a}=\frac{a}{2}=\frac{(Q^2+r^2_{m})^{\frac{3}{2}}r_{m}\sqrt{2}}{2Q^2\sqrt{(Q^2+r^2_{m}+40\pi\rho_{0}r^4_{m}e^{-1}-C_{1}r_{m})(2r^2_{m}-4Q^2)}},
\end{equation}

\begin{equation}
    \bar{b}=\bar{a}\log\left(\frac{2\beta_{m}}{y_m}\right)+b_R-\pi,
\end{equation}

where,
\begin{equation}
    \beta_{m}=\frac{Q^2(Q^2r^2_{m}-2Q^4)}{4r^2_{m}(Q^2+r^2_{m})^2}.
\end{equation}
and
\begin{equation}
    y_m=y(r_{m}).
\end{equation}

Next, the deflection angle, expressed in terms of $\theta$, is given by:

\begin{equation}
    \alpha(\theta)=-\bar{a}\log \left(\frac{\theta D_{ol}}{u_m}-1\right)+\bar{b}.
\end{equation}

\begin{figure}[h]
\includegraphics[width=8.5cm]{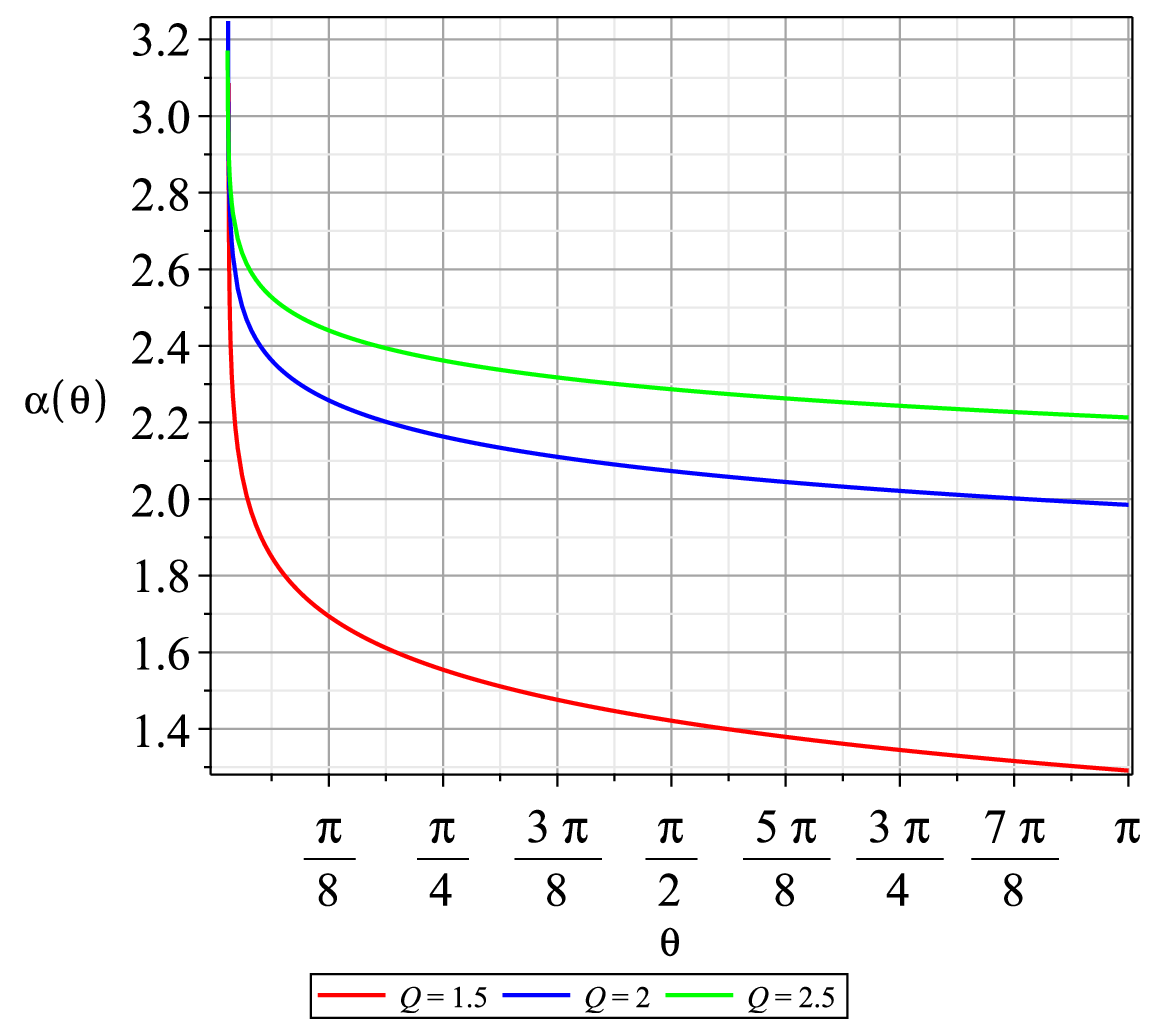}
\includegraphics[width=8.5cm]{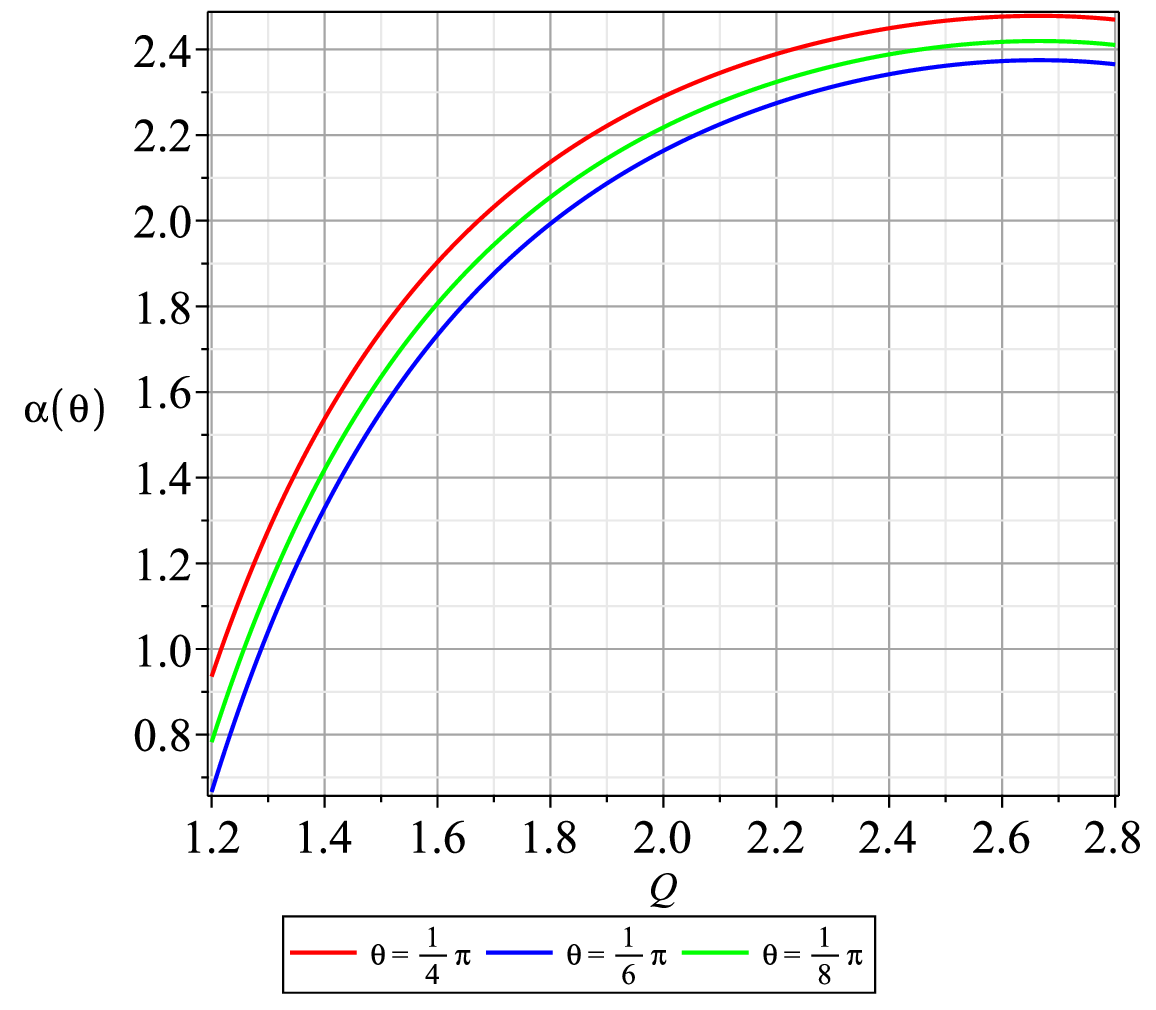}
\caption{ The above diagrams are the graphical  representation of the deflection angle $\alpha(\theta)$ in relation to $\theta$ for various values of $Q$ (left part) and the deflection angle $\alpha(\theta)$ in relation to $Q$ for various values of $\theta$ (right part).} \label{g}
\end{figure}

\newpage
\section{IMAGE ANALYSIS}\label{4}

$D_{ol}$ is previously defined as the constant angle separations among the lens and the viewer. Let $D_{os}$ represent the angular separation among the source and the viewer, and let $D_{ls}$ represent the separation among the source and lens. We consider the line connecting the viewer and lens serve as our axis of reference. The angular gap between the reference axis and the tangent to the null geodesic at the viewer is represented by $\theta$, as was previously indicated.
Let $\beta$ represent the angle among the reference axis and the tangent to the null geodesic at the real location of origin.

$\alpha$ represents the deflection angle. If a lens exhibits an attractive character and produces pictures that are further from the reference axis than from their actual site of origin, the deflection angle is considered to be positive.

Furthermore, it is described as negative if the lens exhibits a nature of repulsiveness, producing pictures that are nearer to the reference axis than they actually are at the source \cite{r60}.

The lens equation has the following expression \cite{r61}:

\begin{equation}
    \sin(\theta-\beta)=\frac{D_{ls}}{D_{os}}\sin\alpha(\theta).
\end{equation}

Reduction of Einstein angle of deflection is defined by

\begin{equation}
    \widehat{\alpha(\theta)}=\sin^{-1}\left(\frac{D_{ls}}{D_{os}}\sin\alpha(\theta)\right).
\end{equation}

We also have,

\begin{equation}
    \sin(\theta)=\frac{u}{D_{ol}},
\end{equation}

i.e,

\begin{equation}
  \sin\theta=\frac{r^2_{0}}{D_{ol}\sqrt{Q^2+r^2_{0}}}.
\end{equation}

The image's magnification is defined by,

\begin{equation}
    \mu=\left(\frac{\sin\beta}{\sin\theta}\frac{d\beta}{d\theta}\right)^{-1}.
\end{equation}

The position of the tangential and radial critical curves can be determined via singularity specifics about the next two expressions:

\begin{equation}
   \mu_{t}=\left(\frac{\sin\beta}{\sin\theta}\right)^{-1},
\end{equation}

\begin{equation}
    \mu_{r}=\left(\frac{d\beta}{d\theta}\right)^{-1}.
\end{equation}

\begin{figure}[h]
\centering
\includegraphics[width=15cm]{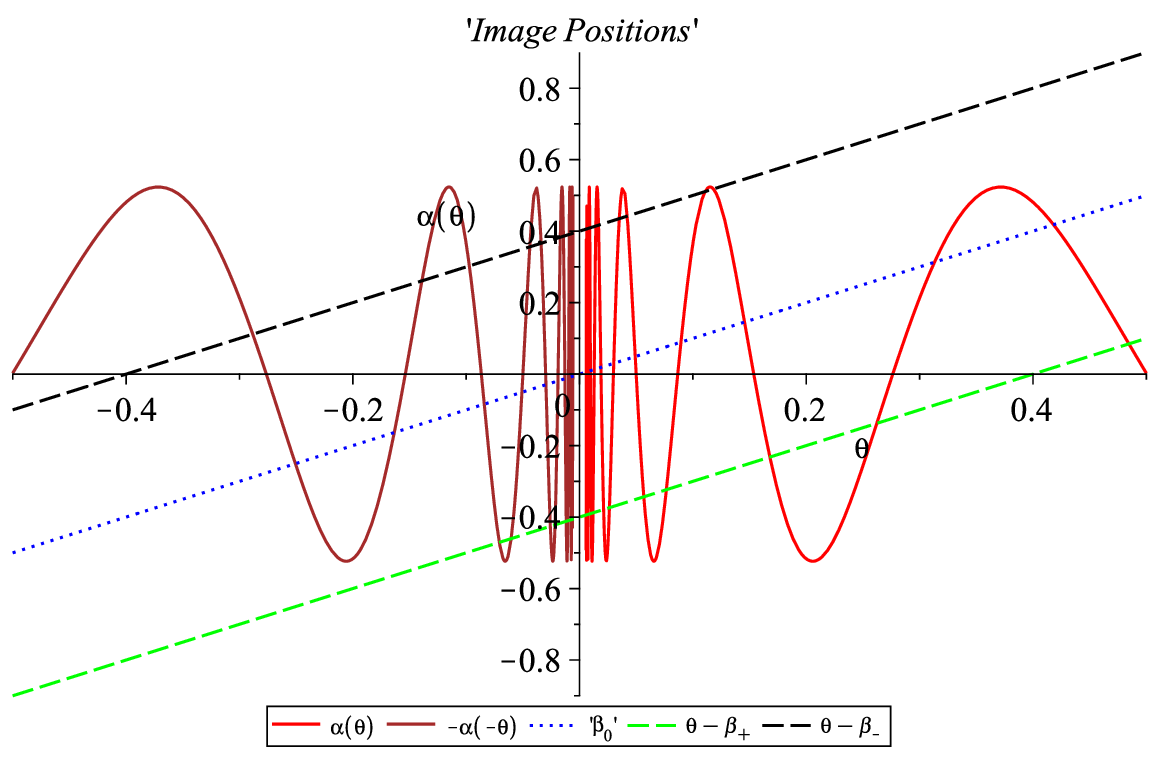}
\caption{The above diagram is the graphical  representation of $\widehat{\alpha}$ and $\theta-\beta$  expressed as $\theta$ dependent and also $-\widehat{\alpha}$ and $-\theta-\beta$  expressed as $-\theta$ dependent, assuming $Q=1$ and $\beta=0$ is illustrated via a direct path with dots that goes via the origin.Also the dashed lines are applied to represent $\theta(\theta)-\beta(\theta)$ and $-\theta(-\theta)-\beta(-\theta)$, assuming $\beta=0.40  arcseconds$} \label{h}
\end{figure}

\begin{figure}[h]
\includegraphics[width=8.5cm]{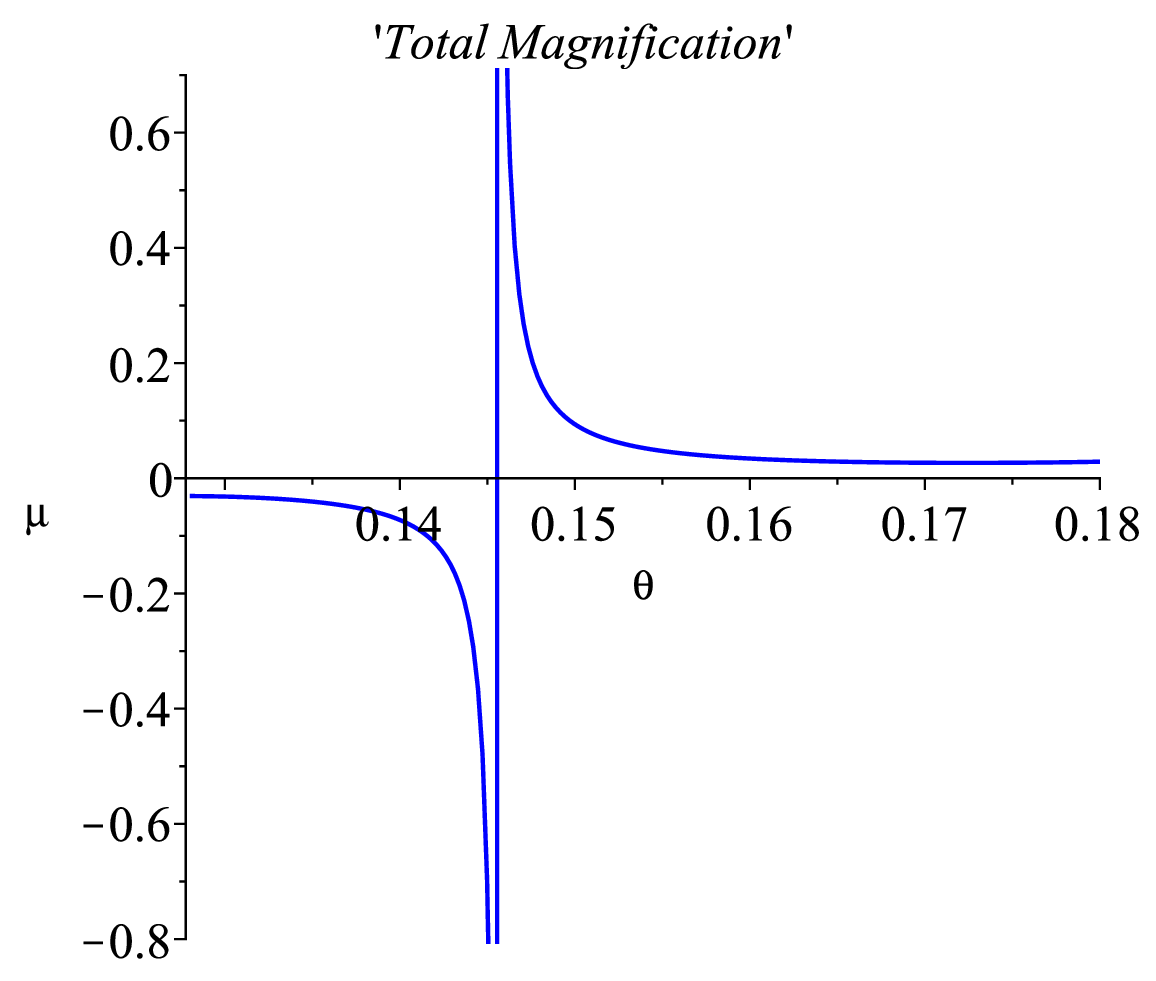}
\includegraphics[width=8.5cm]{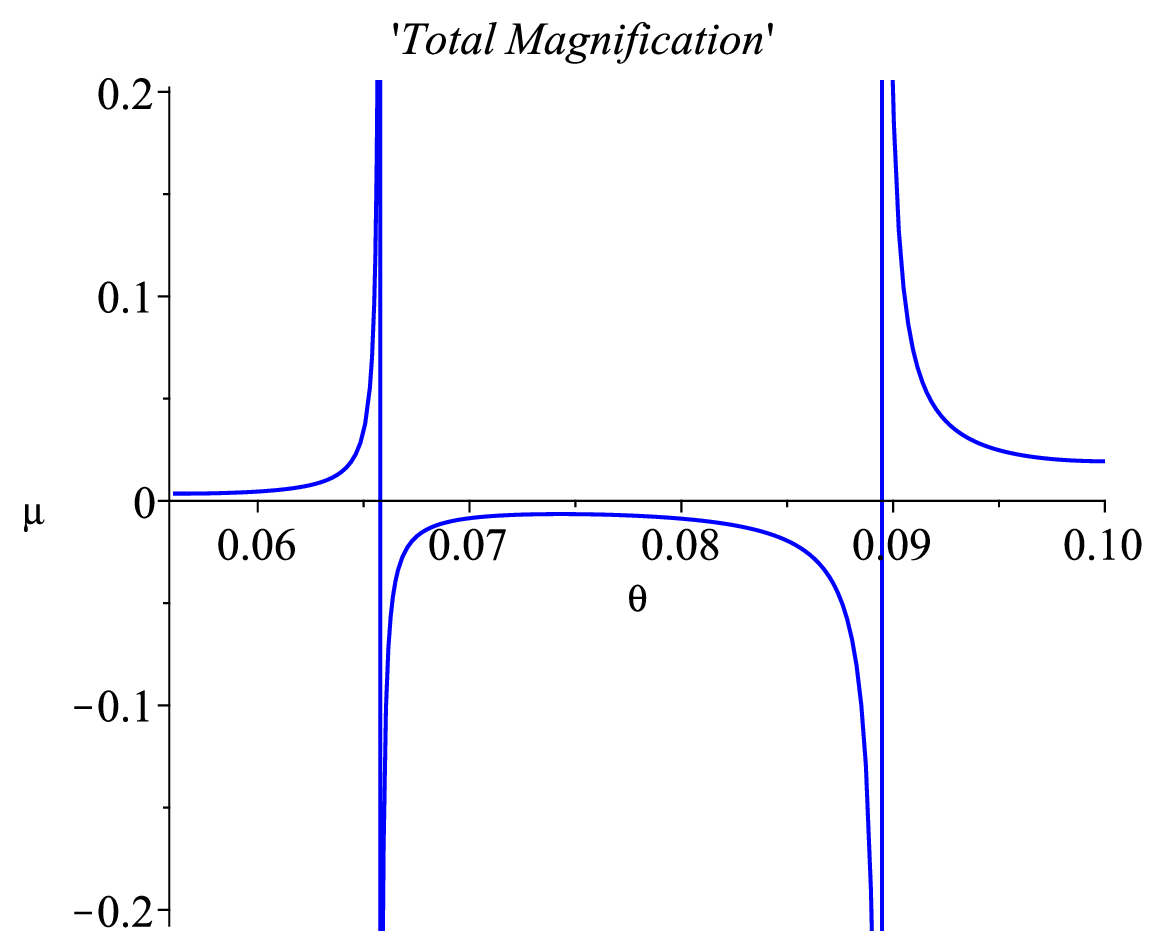}
\caption{The above diagrams are the visual depiction of the total magnification $\mu$.} \label{i}
\end{figure}

\begin{figure}[h]
\includegraphics[width=8.5cm]{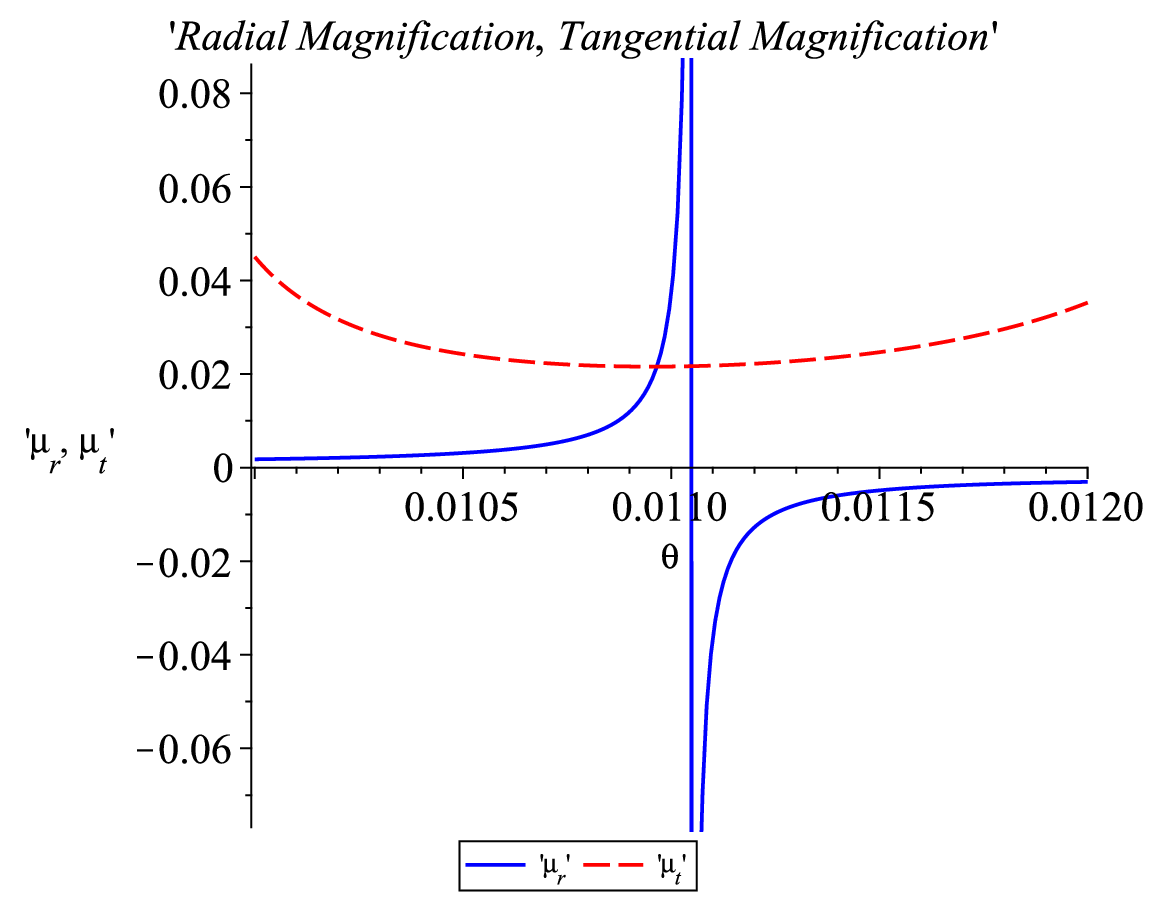}
\includegraphics[width=8.5cm]{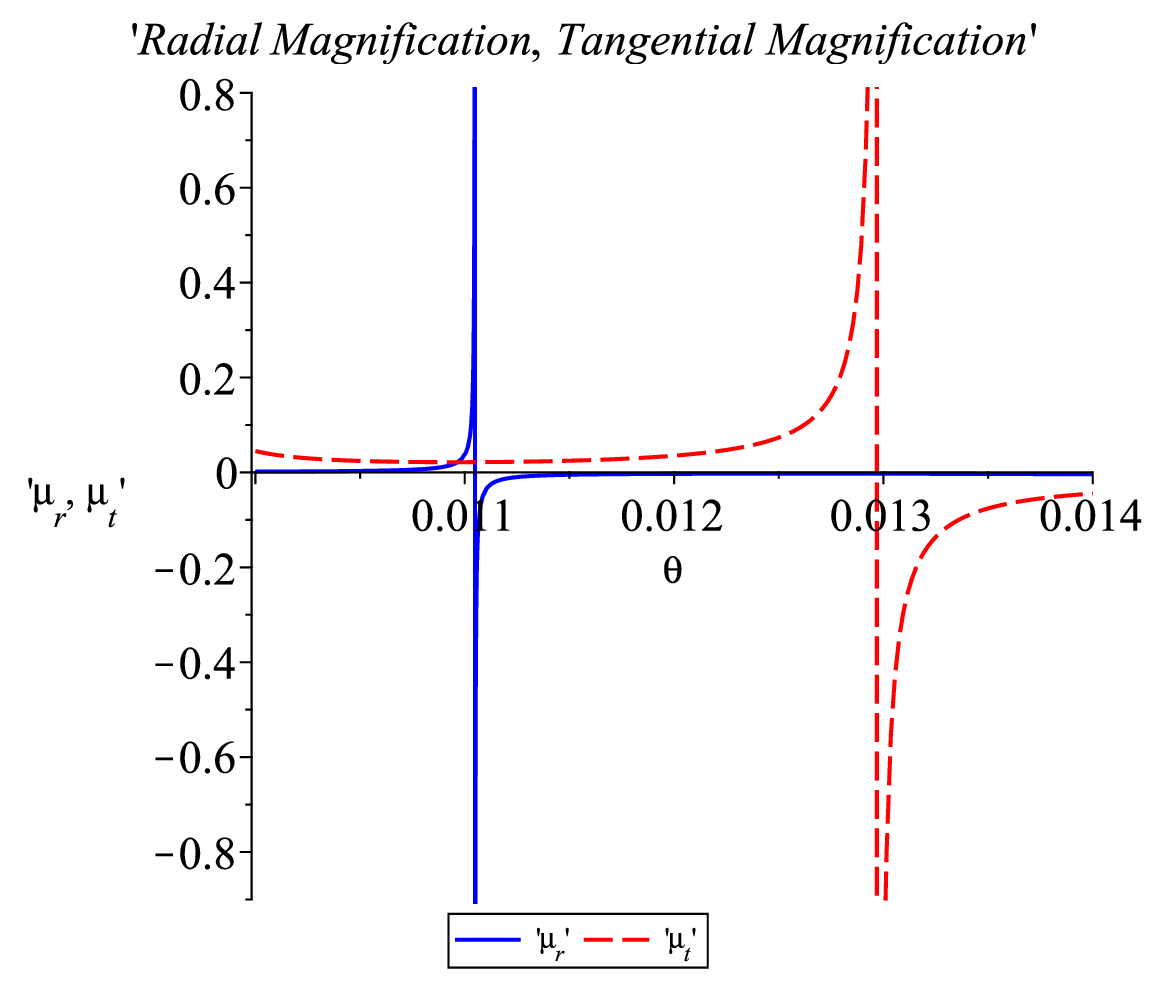}
\includegraphics[width=12cm]{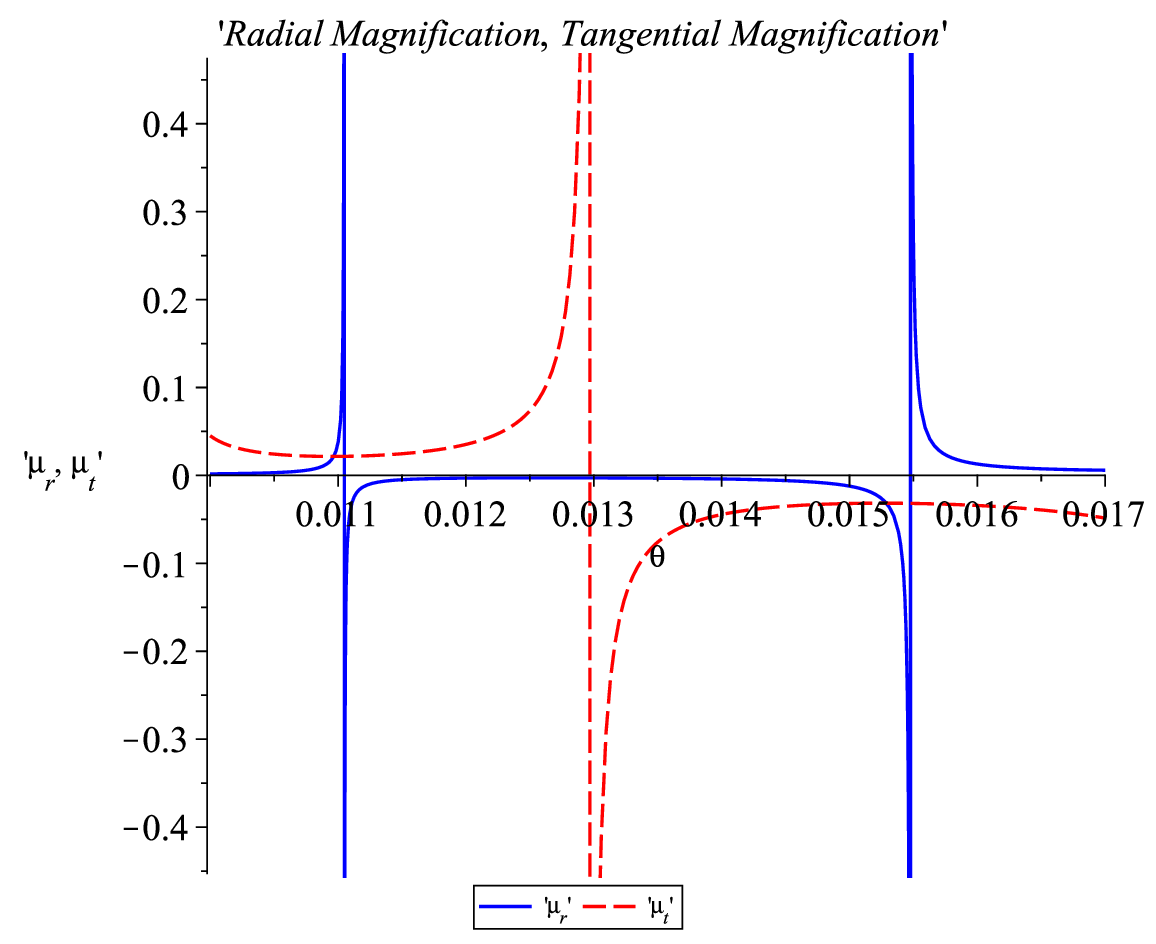}
\caption{The above diagrams are the graphical  representation of the tangential $\mu_{t}$ and radial $\mu_{r}$ magnifications. } \label{j}
\end{figure}

We have drawn the functions $\widehat{\alpha}$ and $\theta-\beta$ expressed as  $\theta$ dependent and also $-\widehat{\alpha}$ and $-\theta-\beta$ expressed as $-\theta$ dependent. Now we are able to analyze the pictures
 on either side regarding the reference axis using this method.Now assumed the ratio $\frac{D_{ls}}{D_{os}}=\frac{1}{2}$.
$\beta$ is indicated via a direct path with dots that goes via the point of origin.
It meets the continuous arcs $\widehat{\alpha(\theta)}$ and$-\widehat{\alpha(-\theta)}$ twice; these points of intersection represent the locations of the Einstein rings, also known as tangential critical curves (TCC).
There have been multiple concentric Einstein rings, as could have been noticed in Fig.(\ref{h}).
Also,the lines have been drawn $\theta(\theta)-\beta(\theta)$ assuming $\beta=0.40 arcseconds$
and identified two the cross-section locations, each has these arcs $\widehat{\alpha(\theta)}$ and$-\widehat{\alpha(-\theta)}$ , accordingly.
The position of the pictures at the points of intersection reveals the pictures on both sides of the reference axis. These two pictures that are nearer to the axis have been described by the term inner image, in addition to pictures that are greater distance are described by the image on the outside. Fig.(\ref{i}) illustrates the whole magnification $\mu$, whereas Fig.(\ref{j}) display the tangential $\mu_{t}$ and radial $\mu_{r}$ magnifications. The exact locations of the radial critical curves (RCC) and tangential critical curves (TCC) have been determined by the positions of singularities of $\mu_{t}$.
The graph illustrates the precise location of a single RCC, and when $\theta$ increases more, the magnification reduces down quickly.

\clearpage

\section{Massive Object's deflection via the Rindler-Ishak Method} \label{5}

The conventional spacetime metric will need to be construct the Jacobi metric, which is necessary for evaluating the deflection of a massive object.

As we have detailed calculations in our previous paper \cite{r56} about this method, we directly state the Jacobi metric as follows:

\begin{equation}
ds^ 2 =  m^2 \left( \frac{1}{1 - v^2} - A \right)  \left[ \frac{B}{A} dr^2 + \frac{C}{A} d\phi^2   \right].
\end{equation}
Consequently, the path of the massive object gets to be stated as
\begin{equation}\label{0o}
\left(\frac{dU}{d\phi} \right)^ 2 = \frac{C^2 U^4}{AB}  \left[\frac{1}{v^2 b^2} -A \left( \frac{1-v^2}{v^2 b^2} + \frac{1}{C} \right) \right]  \equiv f(U).
\end{equation}

When $v$ tends towards 1, Eq.(\ref{0o}) becomes equivalent to the null geodesic equation.

The conditions to obtain a circled orbit as

\begin{figure}[h]
\includegraphics[scale=0.5]{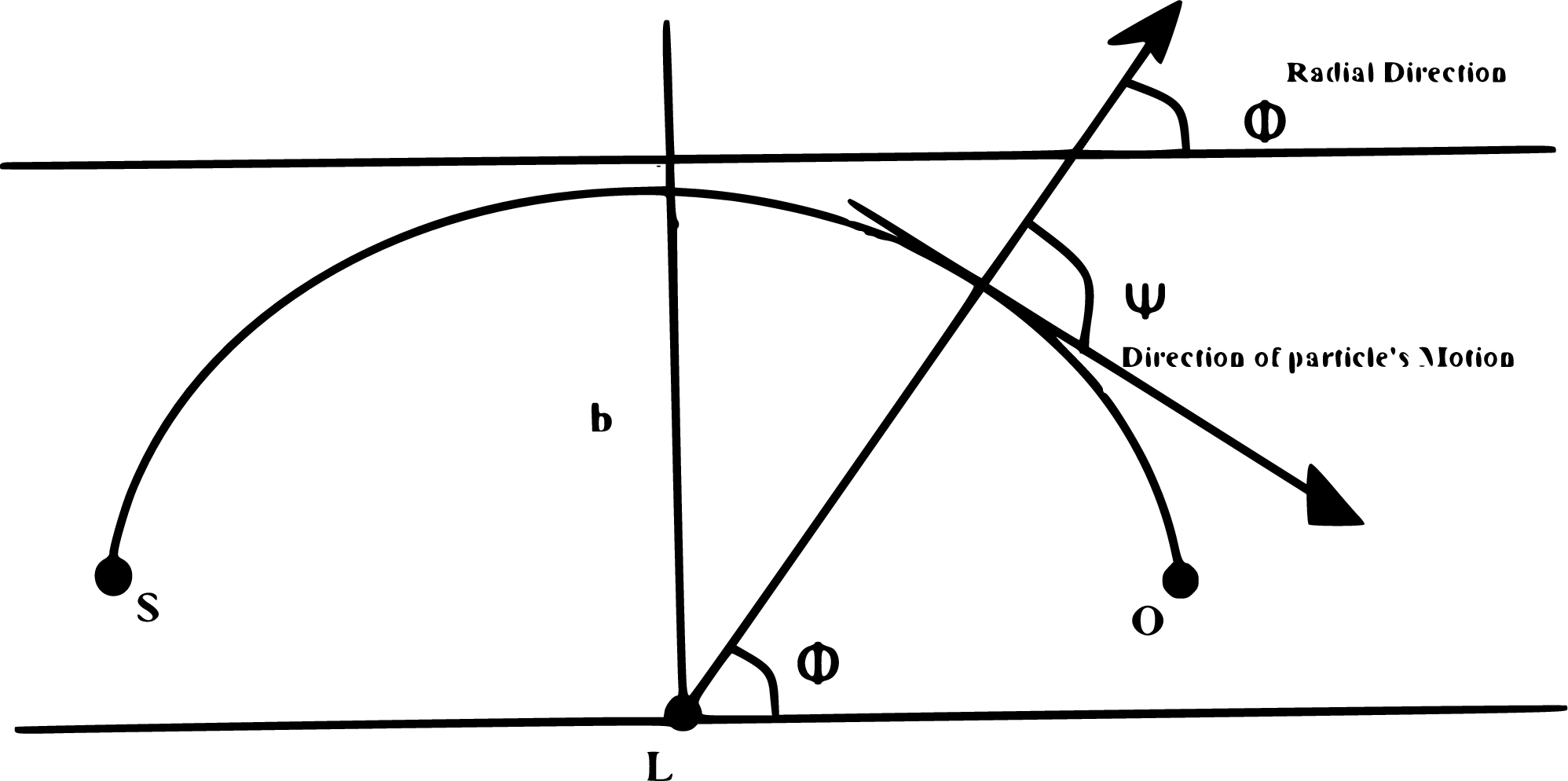}
\caption{The above diagrams provide a visual representation of the movement of a massive object in orbit, especially within the weak field limit, in which the closest approach is nearly the same as the impact parameter, $b$.} \label{o11}
\end{figure}

\begin{equation}
f(U) = 0  \:\:  and  \:\:  f'(U) = 0,
\end{equation}
at some $ U = U_c $.

This equation can be expressed as
\begin{equation}
\frac{d^2 U}{d\phi^2}+U=F(U),\hspace{4mm} where  ~~F(U)=U+\frac{1}{2}{\frac{df}{dU}}.
\end{equation}

The Jacobi Metric is
\begin{equation}\label{e59}
	ds^ 2 =  m^2 \left( \frac{1}{1 - v^2} -(1+\frac{Q^2}{r^2})\right)\left[\left(\frac{r^4}{(Q^2+r^2)\left(8r\pi r_{s}\rho_{s}(r^2+2rr_{s}+2r^2_{s})e^{\frac{-r}{r_{s}}}+Q^2+r^2-rC_{1}\right)}\right)dr^2+\frac{r^4}{Q^2+r^2}d\Omega^2\right].	
\end{equation}
Therefore, the equation for the trajectory takes the form of
\begin{equation}\label{o30}
\left(\frac{dU}{d\phi} \right)^ 2 =-  \left[\frac{U(Q^2b^2U^2v^2+((1-v^2)Q^2+v^2b^2)U^2-v^2}{\left(16\pi(r^2_{s}U^2+r_{s}U+\frac{1}{2})\rho_{s}r_{s}e^{\frac{-1}{Ur_{s}}}+Q^2U^3+U-C_{1}U^2\right)v^2b^2(1+Q^2U^2)}\right].
\end{equation}

This implies
\begin{multline}
  \frac{df}{dU} =-\frac{2\left(\frac{1}{v^2b^2}-(1+Q^2U^2\left(U^2+\frac{1-v^2}{v^2b^2}\right)\right)Q^2U^2}{(1+Q^2U^2)^2\left(1+Q^2U^2-U\left(-8r_{s}(\frac{1}{U^2}+2\frac{2r_s}{U}+2r^2_{s})e^{\frac{-1}{Ur_{s}}}\pi\rho_{s}+C_{1} \right)\right)} \\
  +\frac{2\left(8\pi\rho_{s}(Ur_{s}+1)(U^2r^2_{s}+\frac{1}{2})e^{\frac{-1}{Ur_s}+U^3(Q^2U-\frac{C_1}{2})}\right)(Q^2v^2b^2U^4+((1-v^2)Q^2+v^2b^2)U^2-v^2)}{Uv^2b^2(Q^2U^2+1)\left(16\pi\rho_{s}r_s(U^2r^2_{s}+Ur_s+\frac{1}{2})e^{\frac{-1}{Ur_s}}+Q^2U^3+U-C_{1}U^2\right)^2} \\
  -\frac{2Q^2U\left(U^2+\frac{1-v^2}{v^2b^2}\right)-2(1+Q^2U^+1)U}{{(1+Q^2U^2)\left(1+Q^2U^2-U\left(-8r_{s}(\frac{1}{U^2}+\frac{2r_s}{U}+2r^2_{s})e^{\frac{-1}{Ur_{s}}}\pi\rho_{s}+C_{1} \right)\right)}}.
  \end{multline}

Now,
\begin{equation}\label{o10}
	\frac{d^2 U}{d\phi^2}+U=U+ \frac{1}{2}\frac{df}{dU}.
\end{equation}

Solving the Eq.(\ref{o10}) , We have

\begin{multline}\label{o20}
  U(\phi)=A\sin\phi+B\cos\phi+\frac{1}{336\pi\rho_{s}v^2b^8}\Biggr[-400r^5_{s}v^2\cos^8\phi-72\left(v^2b^2-\frac{244}{9}v^2r^2_{s}+Q^2\right)r^3_{s}\cos^6\phi+\bigg(-210r^4_{s}v^2b\sin\phi \\-4160r^5_sv^2+(192v^2b62+316Q^2)r^3_{s}\bigg)\cos^4\phi+\bigg(945v^br^4_{s}\sin\phi+6880v^2r^5_{s}+(-72v^2b^2-1080Q^2)r^3_{s}\bigg)\cos^2\phi \\-168\phi\bigg(\pi b^6\rho_{s}+\frac{75r^4_{s}}{8}\bigg)bv^2\cos\phi+168bv^2(\pi b^6\rho_{s}+5r^4_{s})\sin\phi-360\bigg(-\frac{2}{5}v^2b^2-\frac{34}{9}v^2r^2_{s}+Q^2\bigg)r^3_{s}\Biggr]+O\left(\frac{1}{b^9}\right).
\end{multline}

\begin{figure}[h]
\includegraphics[width=8.0cm]{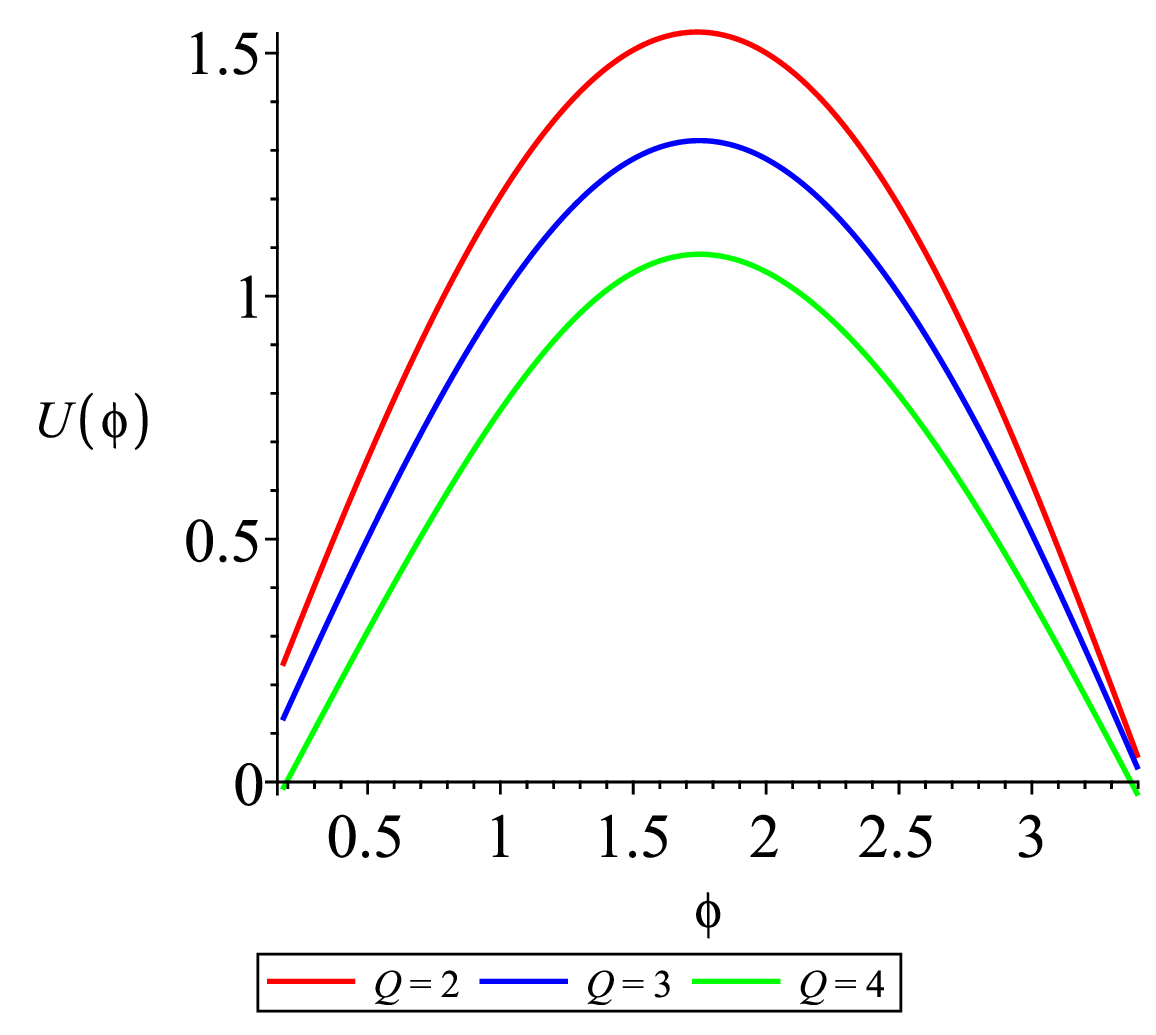}
\includegraphics[width=8.2cm]{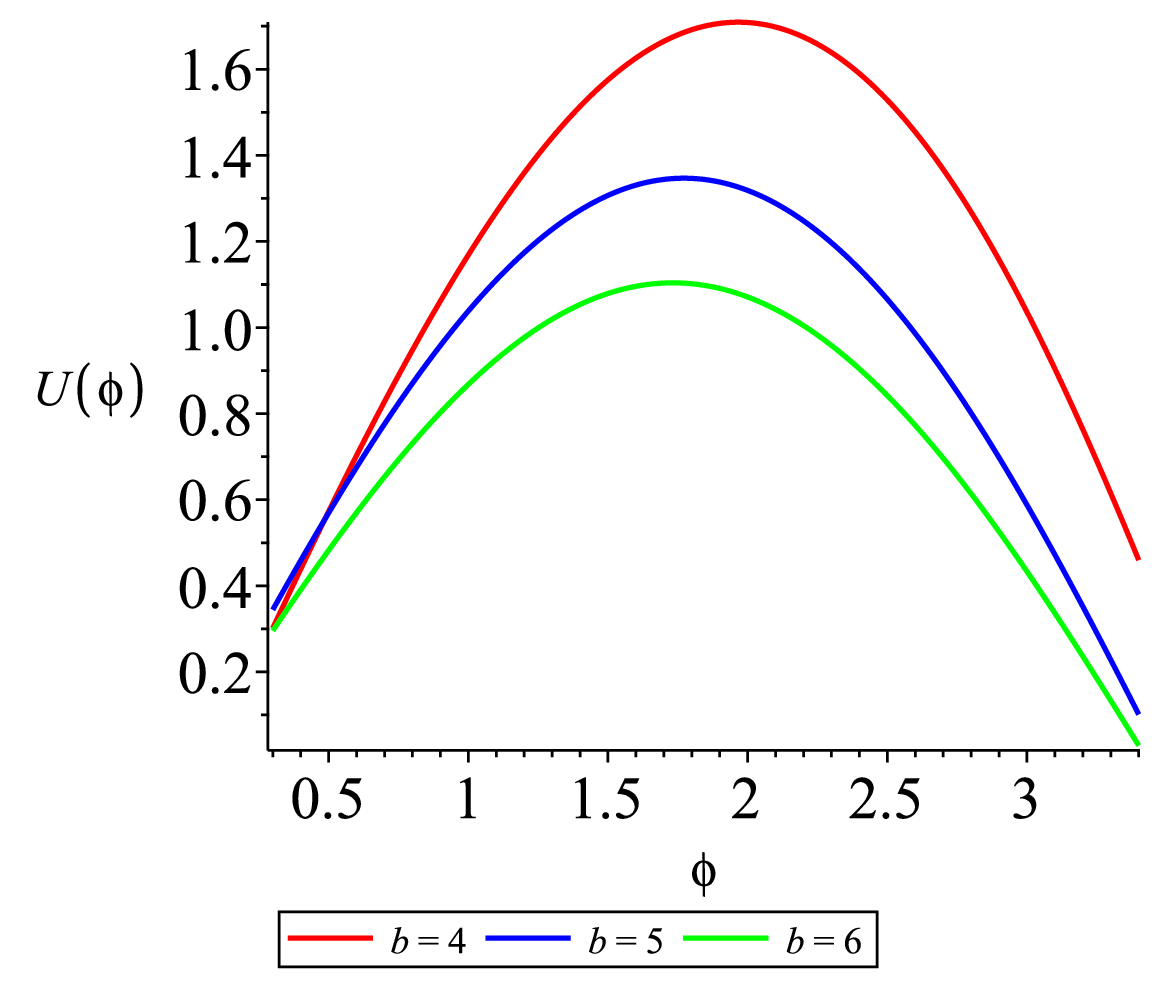}
\caption{ The above diagrams are the graphical  representation of the trajectory $U(\phi)$ as $\phi$ dependent with varying values of $Q$ (left part) and for various values of $b$(right part). }
\end{figure}

Our goal is to resolve this equation utilizing conventional perturbation theory. Using the Rindler-Ishak approach \cite{r62,r63}, the deflection angle of massive particles might be derived. As shown in Fig.(\ref{o11}), the angle among the radial direction ($x^i$) and the particle's motion direction ($y^i$) is represented via $\psi$, in which
\begin{equation} \label{20}
\cos \psi=\frac{g_{ij}x^iy^j}{\sqrt{g_{ij}x^ix^j \sqrt{g_{ij}y^iy^j}}}.
\end{equation}
Here, the components of $g_{ij}$ are provided in Eq.(\ref{e59}).We take $t$ to be constant and $\theta=\frac{\pi}{2}$.

It is important to notice it \[ x^i=(\gamma,1)d\phi,~
and~ y^i=(1,0)dr,\] where, $\gamma=\frac{dr}{d\phi}$.

Currently, Eq.(\ref{20}) is expressed as
\begin{equation} \label{21}
\cos \psi = \frac{|\gamma|}{\sqrt{\gamma^2 + \frac{g_{\phi \phi}}{g_{r r}}}} ~or~ \tan \psi =\frac{ \sqrt{ \frac{g_{\phi \phi}}{g_{r r}}} } {|\gamma|}.
\end{equation}
As shown in Fig. (\ref{o11}), it is noticeable that aside of the deflection angle is expressed by \begin{equation} \label{22}
\alpha^{RI} = 2\psi(\phi)-2\phi .\end{equation} \\
Hence    $\alpha^{RI}$ represents the overall deflection angle.

By applying Eq. (\ref{e59}) and Eqs. (\ref{o20}-\ref{22}), we obtain the overall deflection angle for a massive object,

\begin{equation}
    \alpha^{RI}=2\arctan\left(\frac{\left(\frac{g_{\phi\phi}}{g_{rr}}\right)^{\frac{1}{2}}}{\left|-\frac{1}{U^{2}(\phi)}\frac{dU(\phi)}{d\phi}\right|}\right)-2\phi .
\end{equation}

\begin{figure}[h]
\includegraphics[scale=0.3]{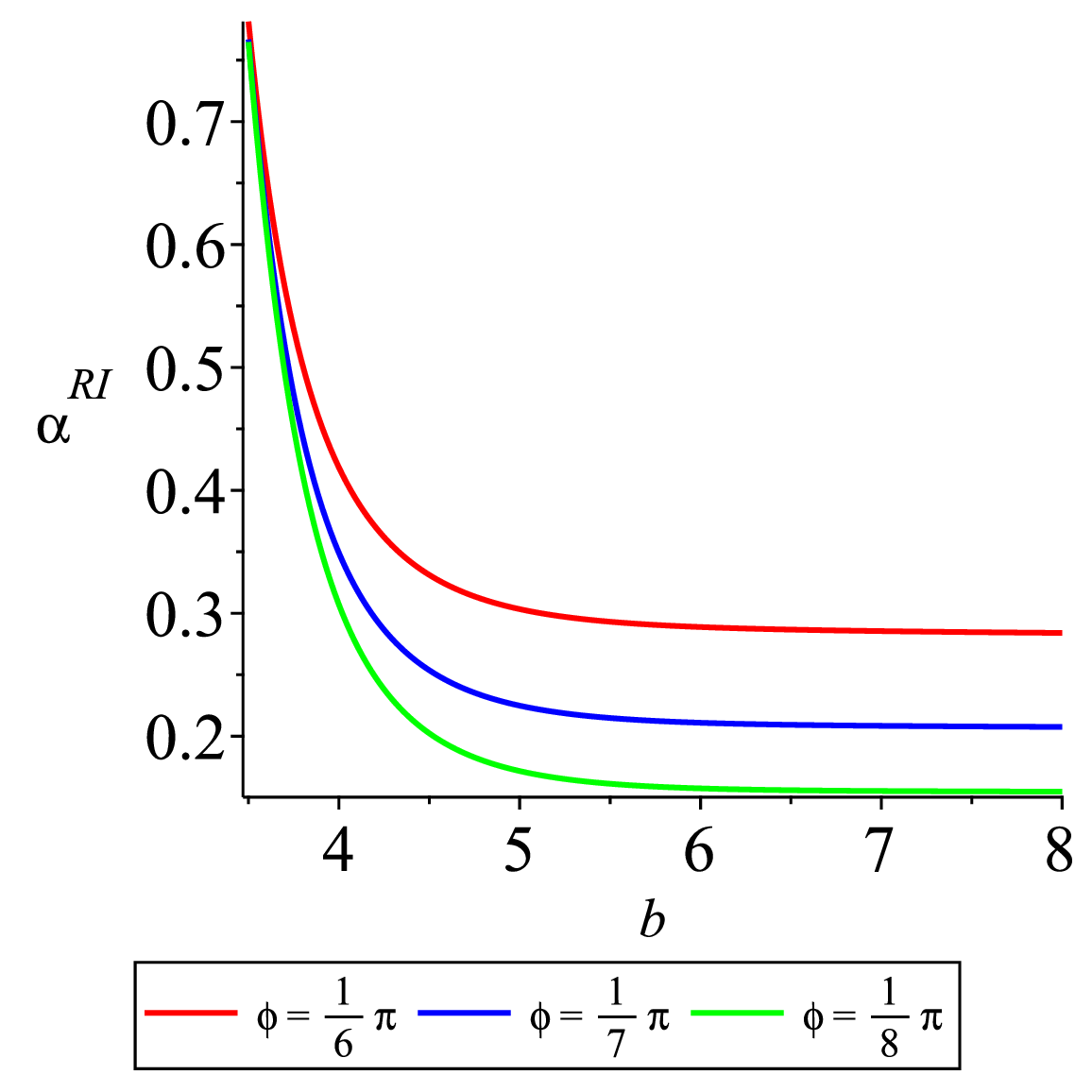}
\includegraphics[scale=0.3]{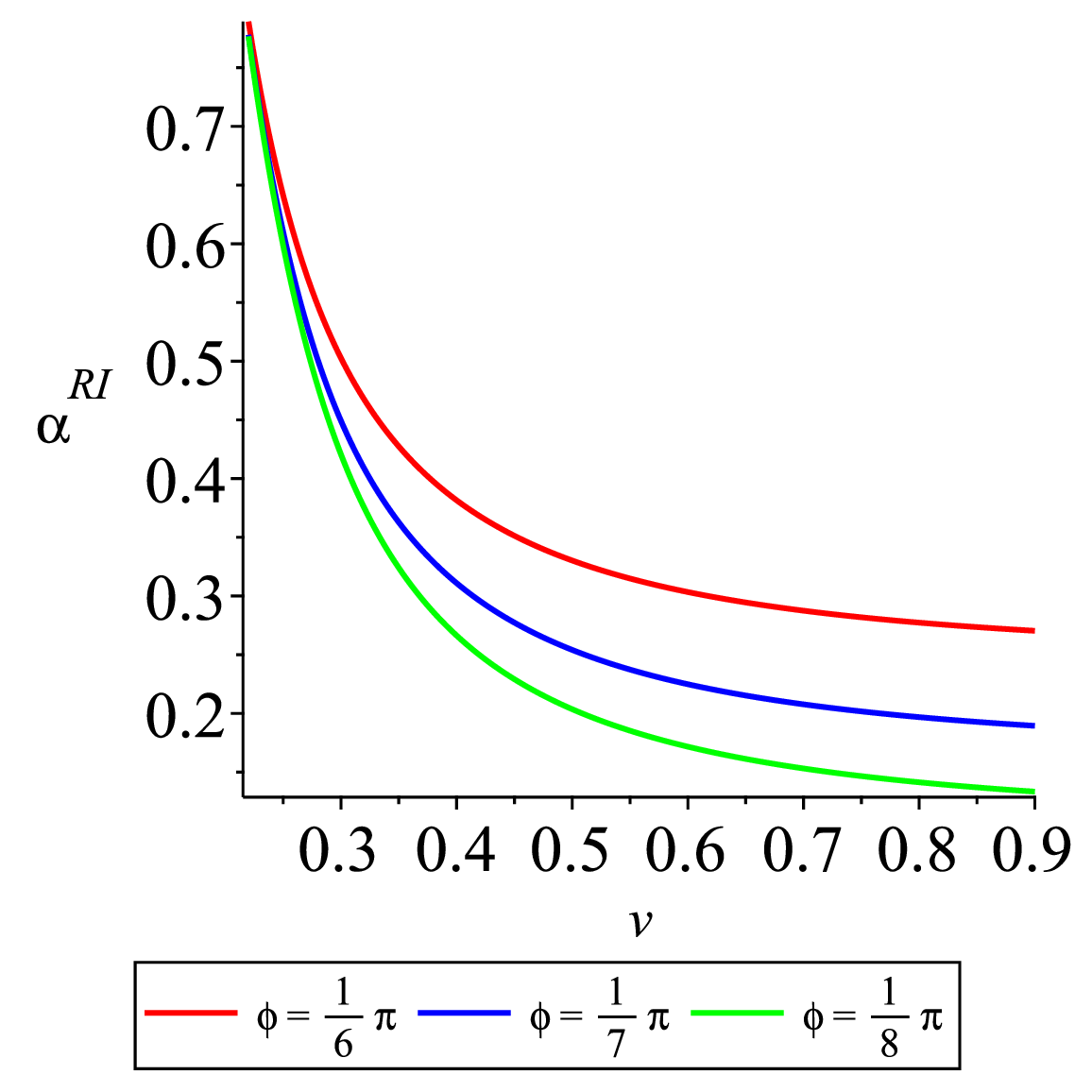}
\includegraphics[scale=0.3]{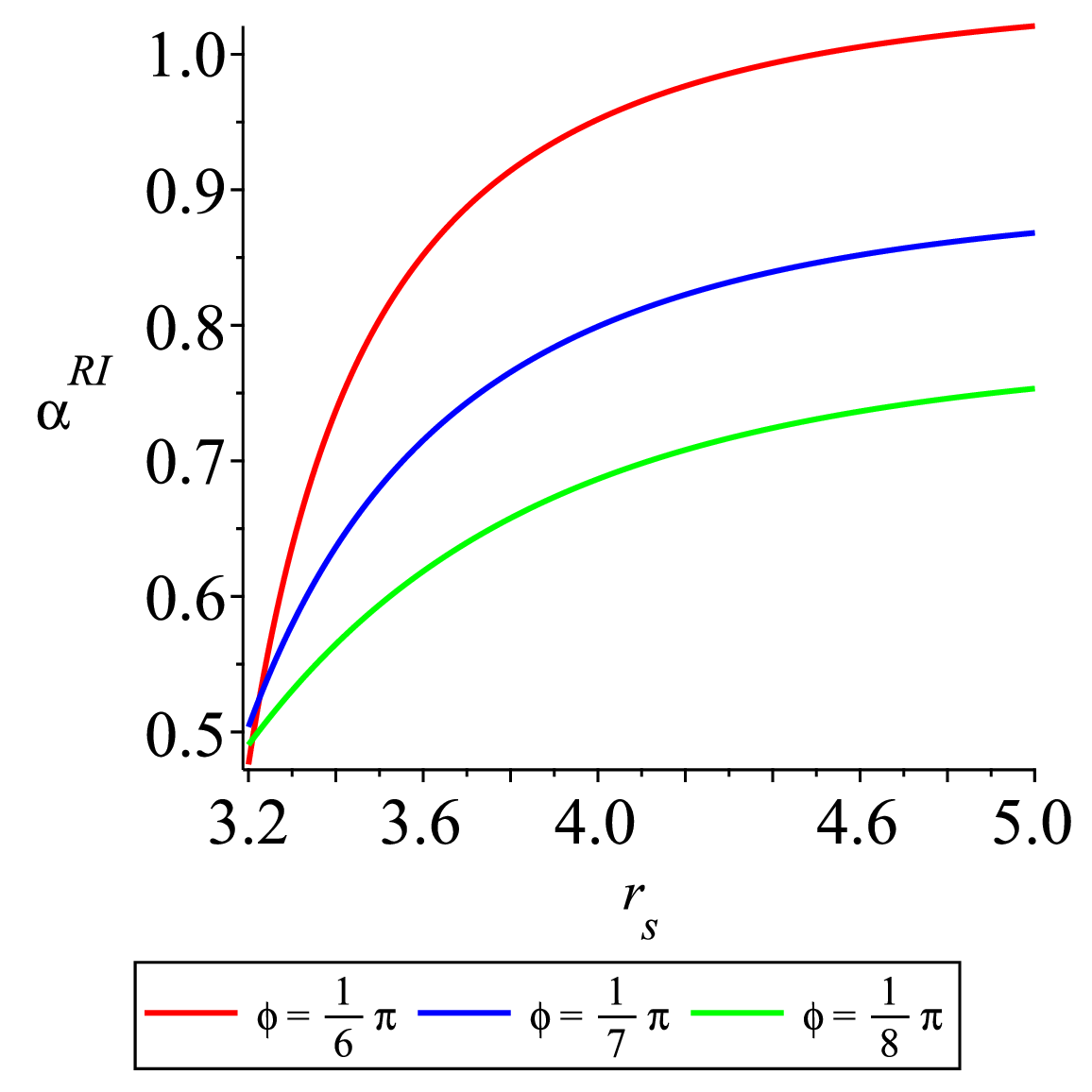}
\caption{The above diagrams are the visual  representation of the deflection angle $\alpha^{RI}$ as $b$ dependent (left part), $v $ (middle part)  and $r_s$ (right part) for various values of $\phi$. Here we assume that $Q=2$.}
\end{figure}

\pagebreak

\section{Massive Object's deflection via the Gauss-Bonnet method} \label{6}

This method, depending on the Gauss-Bonnet theorem, is particularly interesting since the deflection of the particle is independent of coordinates. Using the Jacobi metric $J_{ij}$, which is developed from the Riemannian spacetime, one can define an area on the Jacobi reference geometry and find the deflection.Given the potential to be interpreted as a topological impact, this technique is particularly intriguing. In actuality, a connection between the topology of the manifold and the curvature of a Riemannian space is established via the Gauss-Bonnet.
As you can see, the Jacobi metric includes the arc length parameter. This will be followed by a utilization of the Gauss-Bonnet theorem in a $2D$ Riemannian space with a Jacobi reference space represented via $\mathcal{M}^J$.

\begin{figure} [thbp]
\centering
	\includegraphics[width=8cm]{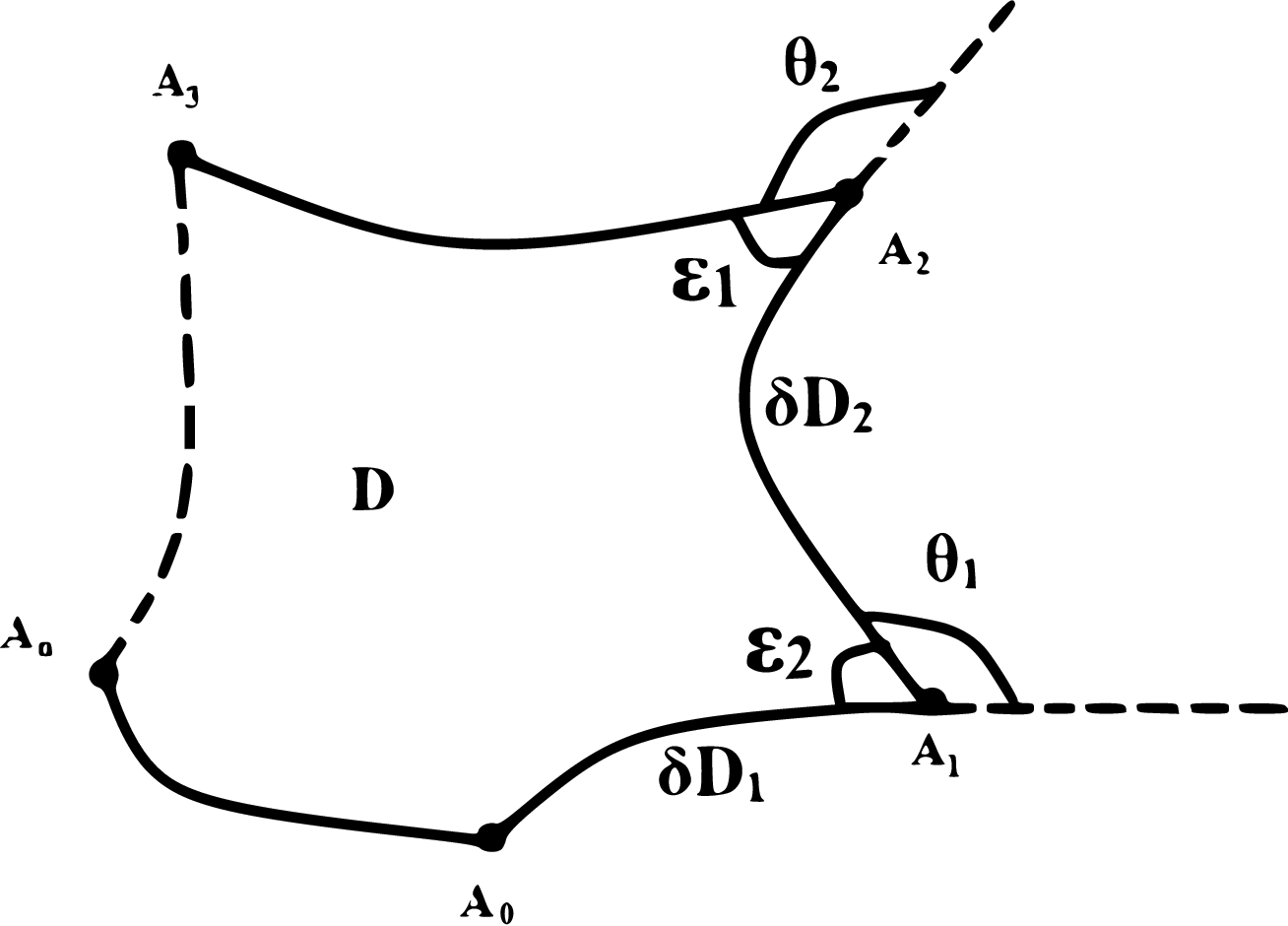}
\includegraphics[width=8cm]{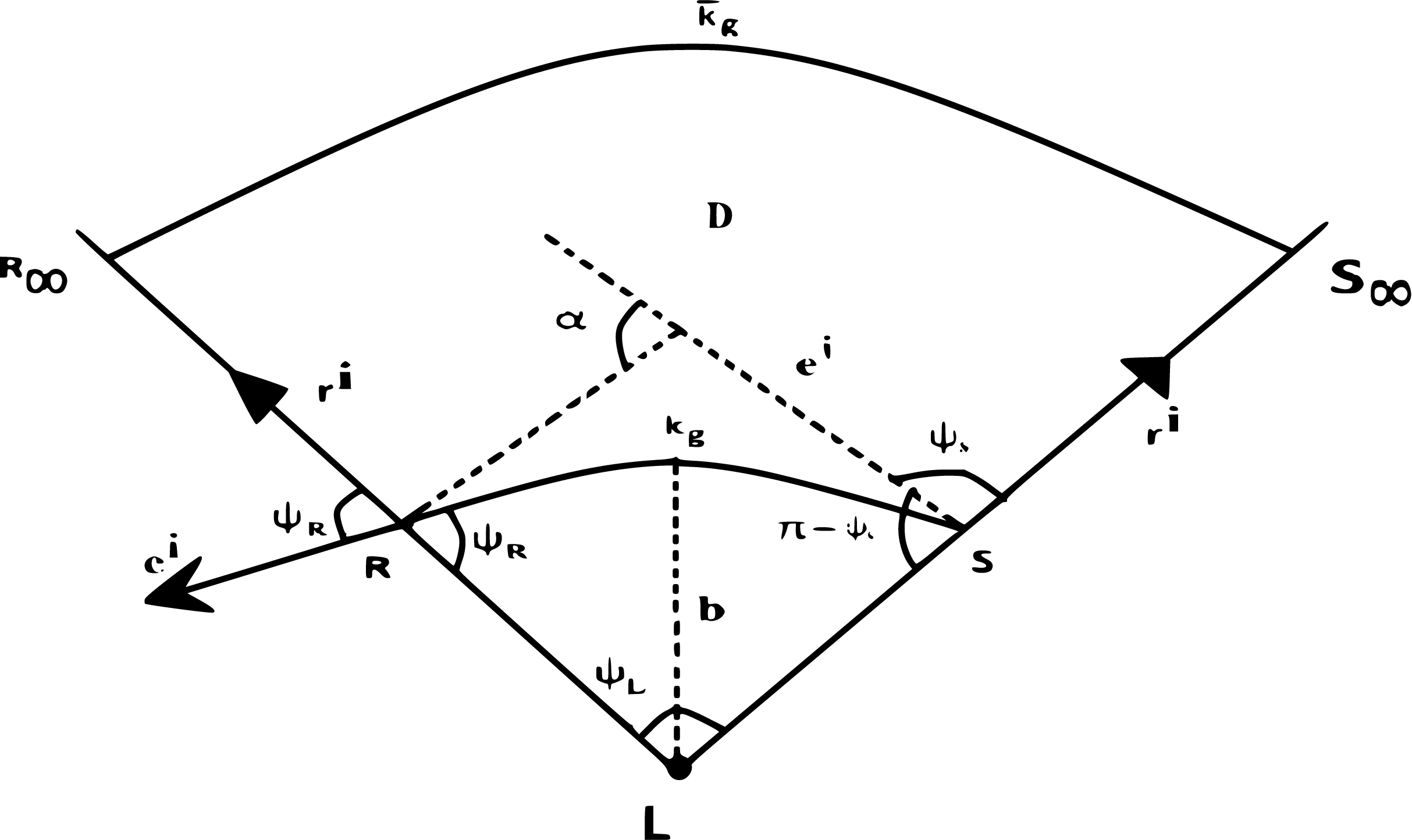}
	\caption { The diagram on the left illustrates the Gauss-Bonnet theorem:  The area $D$ is completely enclosed with the curves $\partial D = \cup_i \partial D_i$, in which the internal angle $\epsilon_i$ and the angle formed outside (jump) $\theta_i$ at the $i^{th}$ vertex in the counterclockwise sense. The diagram on the right illustrates the Gauss-Bonnet theorem in the context of Riemann-Jacobi geometry.} \label{F5}
\end{figure}

As we have detailed calculations in our previous paper \cite{r56} about this method, we directly state the Gaussian curvature as follows:

 \begin{equation} \label{37}
 K =- \frac{1}{\sqrt{J_{rr}} \sqrt{J_{\phi \phi}}}\left[\frac{\partial}{\partial r}\left(\frac{1}{\sqrt{J_{rr}}}\frac{\partial}{\partial r}\sqrt{J_{\phi \phi}}\right)\right]
 - \frac{1}{\sqrt{J_{rr}} \sqrt{J_{\phi \phi}}}\left[  \frac{\partial}{\partial \phi}\left(\frac{1}{\sqrt{J_{\phi \phi}}}\frac{\partial}{\partial \phi}\sqrt{J_{r r}} \right) \right].
\end{equation}

\noindent The deflection angle might be determined by evaluating the above $K$, provided that Riemann-Jacobi geometry is asymptotically Euclidean.

\begin{equation} \label{38}
\alpha =    - \int_{0}^{\pi} \int_{r_0}^{\infty} K \sqrt{detJ} \,dr \,d\phi.
\end{equation}

Using the standard perturbation approach, we consider that the solution of Eq.(\ref{o30}) takes the form $r \approx r_0 +  r_1$, where $  r_0  = \frac{b}{\sin \phi}$ represents the zeroth-order particle trajectory and $r_1$ is the first-order correction to the trajectory.

Therefore, the deflection angle at first order is able to be described as
\begin{equation} \label{39}
\alpha^{GB} \approx    - \int_{0}^{\phi} \int_{\frac{b}{\sin \phi}}^{\infty} K \sqrt{detJ} \,dr \,d\phi.
\end{equation}

Due to complicated form of this integral, we will discuss
the nature of $\alpha^{GB}$ graphically.

\begin{figure}[h]
\includegraphics[scale=0.3]{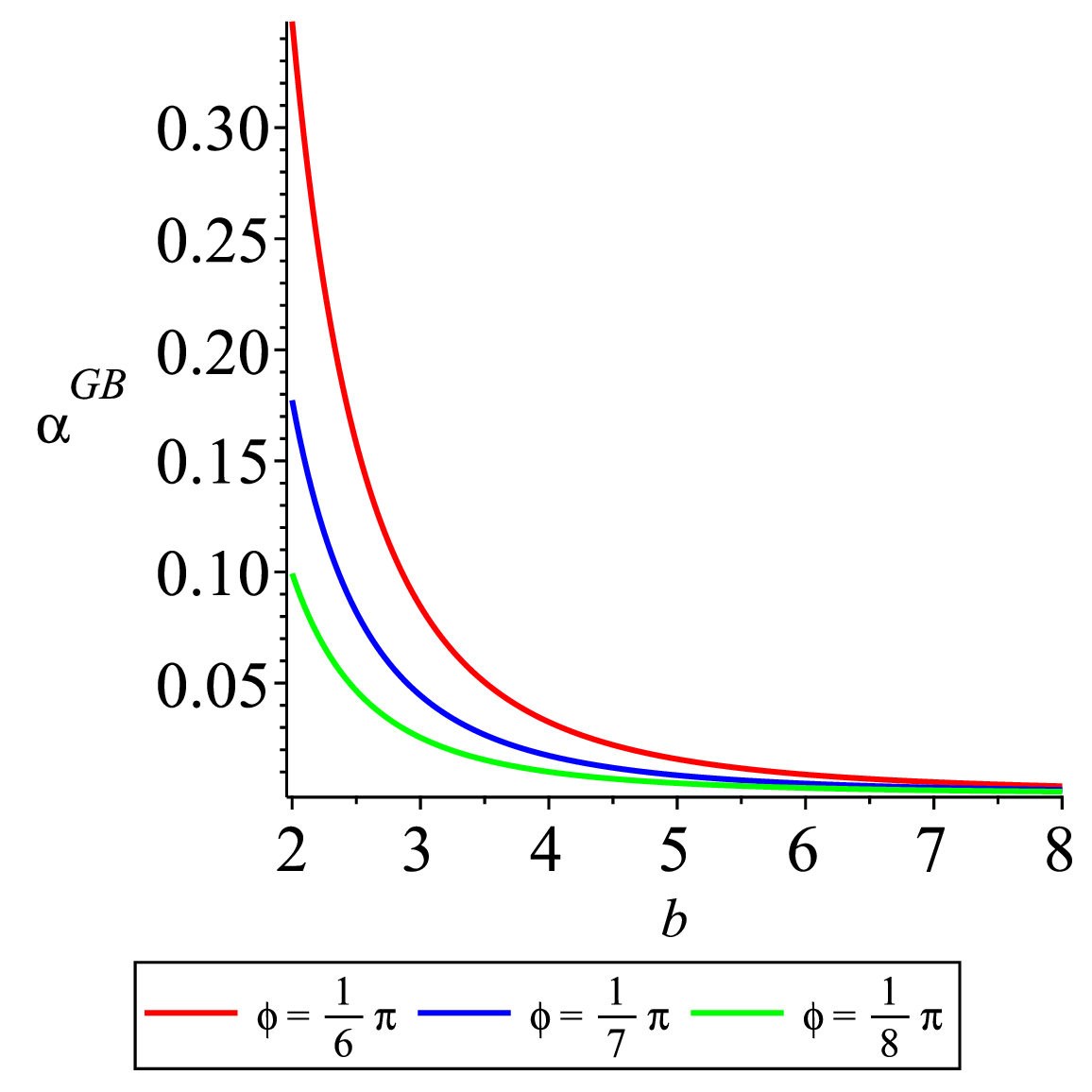}
\includegraphics[scale=0.3]{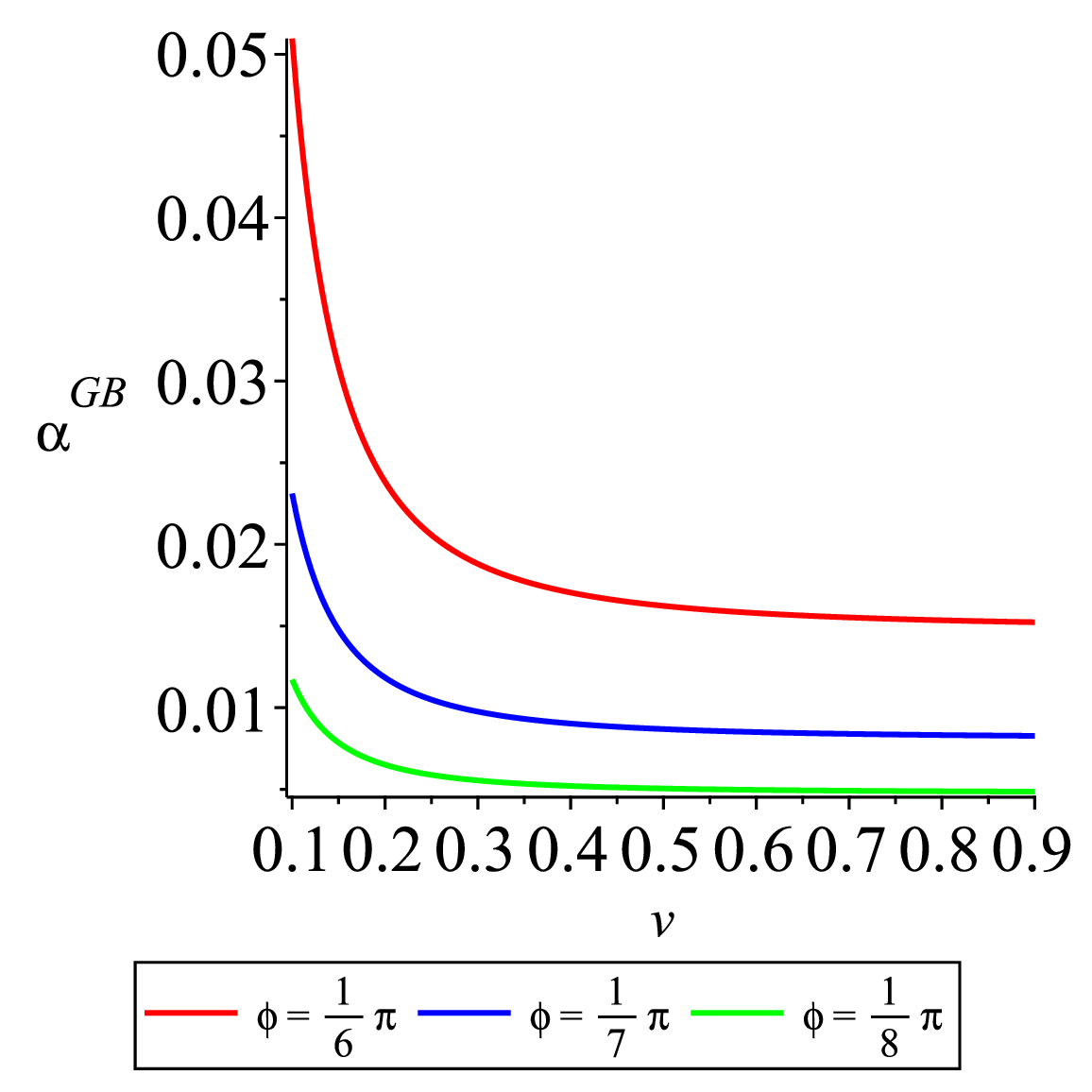}
\includegraphics[scale=0.3]{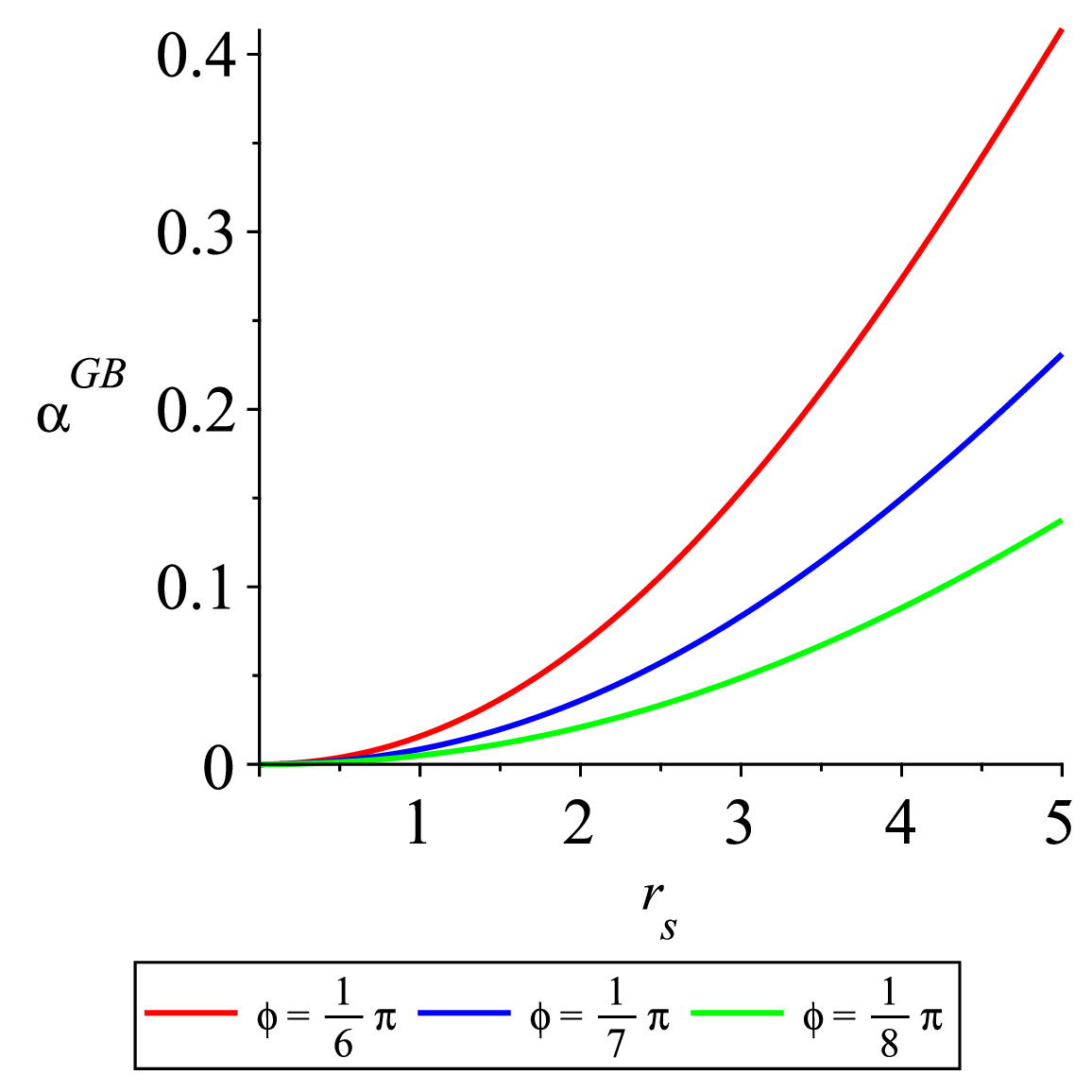}
\caption{The above diagrams are the visual  representation of the deflection angle $\alpha^{GB}$ as functions of parameter $b$ (left part), $v $ (middle part)  and $r_s$ (right part)  for various values of $\phi$.Here we assume that $Q=2$.}.
\end{figure}

\pagebreak

\section{CONCLUSION} \label{7}
\begin{itemize}

\item Photons from a faraway light source are deflected as they move near the wormhole, as predicted by general relativity. By calculating the deflection angle and other crucial image characteristics, we have examined the angular variation between the visible image and the real location of the lensing object (charged galactic wormhole). The analysis of lensing coefficients, including $\bar{a}$, $\bar{b}$, and $\bar{u_m}$, as functions of the parameter $Q$, reveals key trends in the deflection angle of light. As shown in Fig. (\ref{f}), the deflection angle $\alpha(r_0)$ decreases as $r_0$ increases. Furthermore, Fig. (\ref{g}) demonstrates that the deflection angle $\alpha(\theta)$ decreases with increasing $\theta$, while it increases with increasing $Q$, as anticipated. In Fig. (\ref{h}), multiple concentric Einstein rings are observed, highlighting the complex nature of gravitational lensing in this scenario. Finally, Fig. (\ref{i}) illustrates the singularity of radial magnifications ($\mu_r$), which result in several tangential critical curves, two radial critical curves, or the singularity of tangential magnifications ($\mu_t$), providing insight into the intricate structure of gravitational lensing in this context.

\item  The deflection angle $\alpha$ of a massive object around the wormhole exhibits a consistent behavior across both the Gauss-Bonnet and Rindler-Ishak methods. Specifically, $\alpha$ decreases as the impact parameter $b$ increases and also decreases as the velocity $v$ increases. It is observed that particles with lower velocities exhibit less deviation in the GB method compared to the RI method. But, particles with higher velocities show less deviation in the RI method compared to the GB method. A striking and intriguing observation emerges from the analysis: as the velocity approaches the speed of light ($v \to 1$), the deflection angle $\alpha$ not only increases with the scale radius $r_s$, but this behavior mirrors the outcomes of strong gravitational lensing in both the Gauss-Bonnet and Rindler-Ishak methods. This convergence underscores the profound connection between high-velocity particle motion around a wormhole and the lensing effects typically seen in extreme gravitational fields, revealing a fascinating parallel to strong lensing phenomena. This highlights the intricate relationship between the deflection angle, velocity, and the scale radius, providing insightful information about the gravitational dynamics of objects near wormholes.

\item An interesting fact is that the GB method is identical for certain values of $r_s$, as v approaches 1 in the strong field limit. However, the RI method shows some deficiencies. Therefore, it appears that the GB method provides a more accurate prediction of the deflection of both massive and massless particles.

\end{itemize}

Our research provides insightful information about the application of gravitational lensing for understanding galaxies with dark matter density profiles. As advanced technologies such as next-generation space telescopes and more precise gravitational wave detectors become available, our findings will serve as a foundation for more accurate and high-resolution observations of lensing effects. These advancements will enable astronomers to map dark matter distributions in even the faintest and most distant galaxies, offering the potential to probe smaller-scale lensing phenomena and better understand the interaction between dark matter halos and visible matter \cite{r64,r65,r66}. By refining dark matter density profiles, such as the Navarro-Frenk-White (NFW) model, and analyzing lensing distortions, our work aids in constraining dark matter models and testing alternative theories of gravity. As gravitational lensing remains a powerful tool for mapping mass distributions, our research contributes to advancing the understanding of galaxy dynamics, rotation curves, and the formation of galactic structures, which will ultimately enhance our comprehension of the role of dark matter in the creation of galaxies and the enormous scale structure of the universe.

\begin{figure}[h]
\includegraphics[scale=0.3]{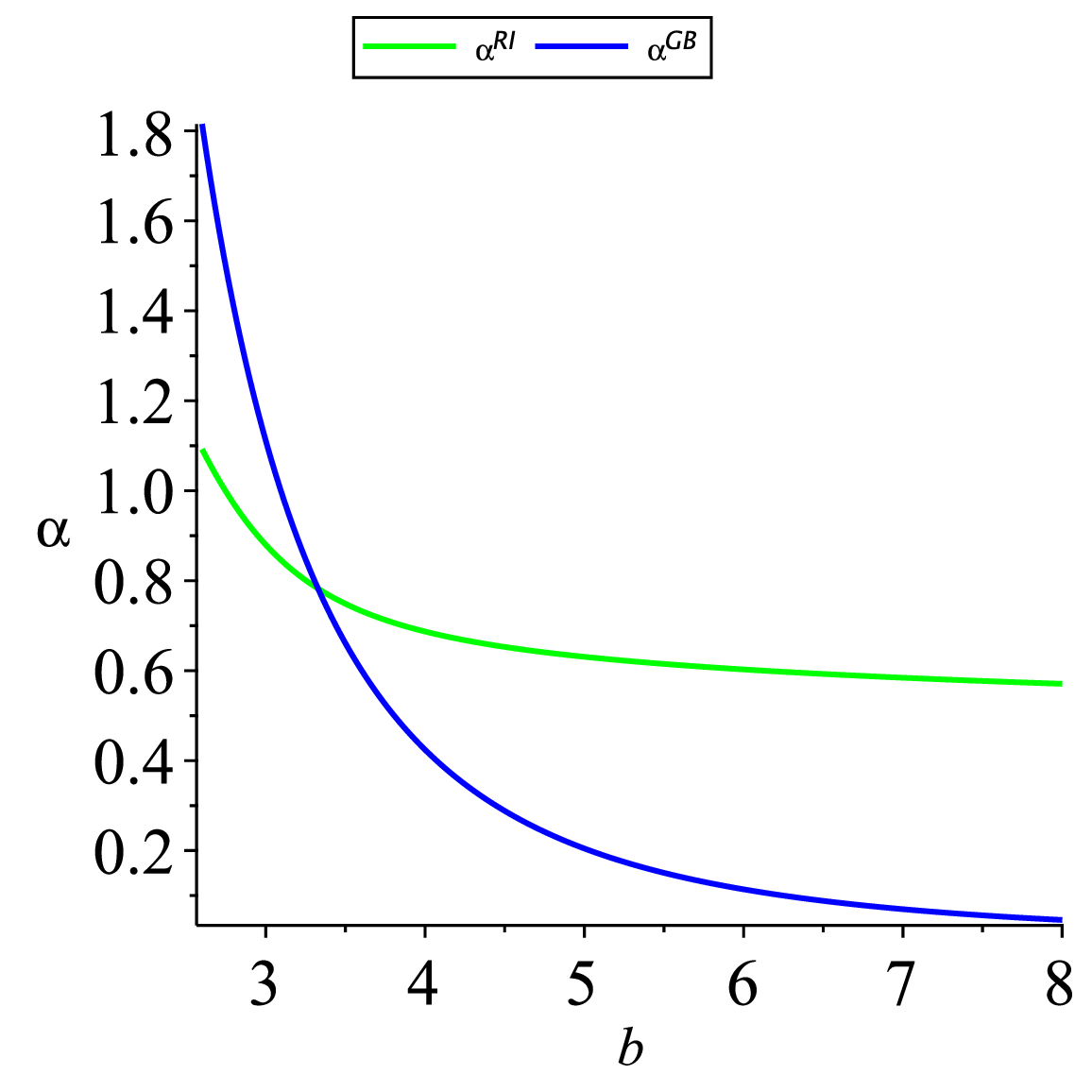}
\includegraphics[scale=0.3]{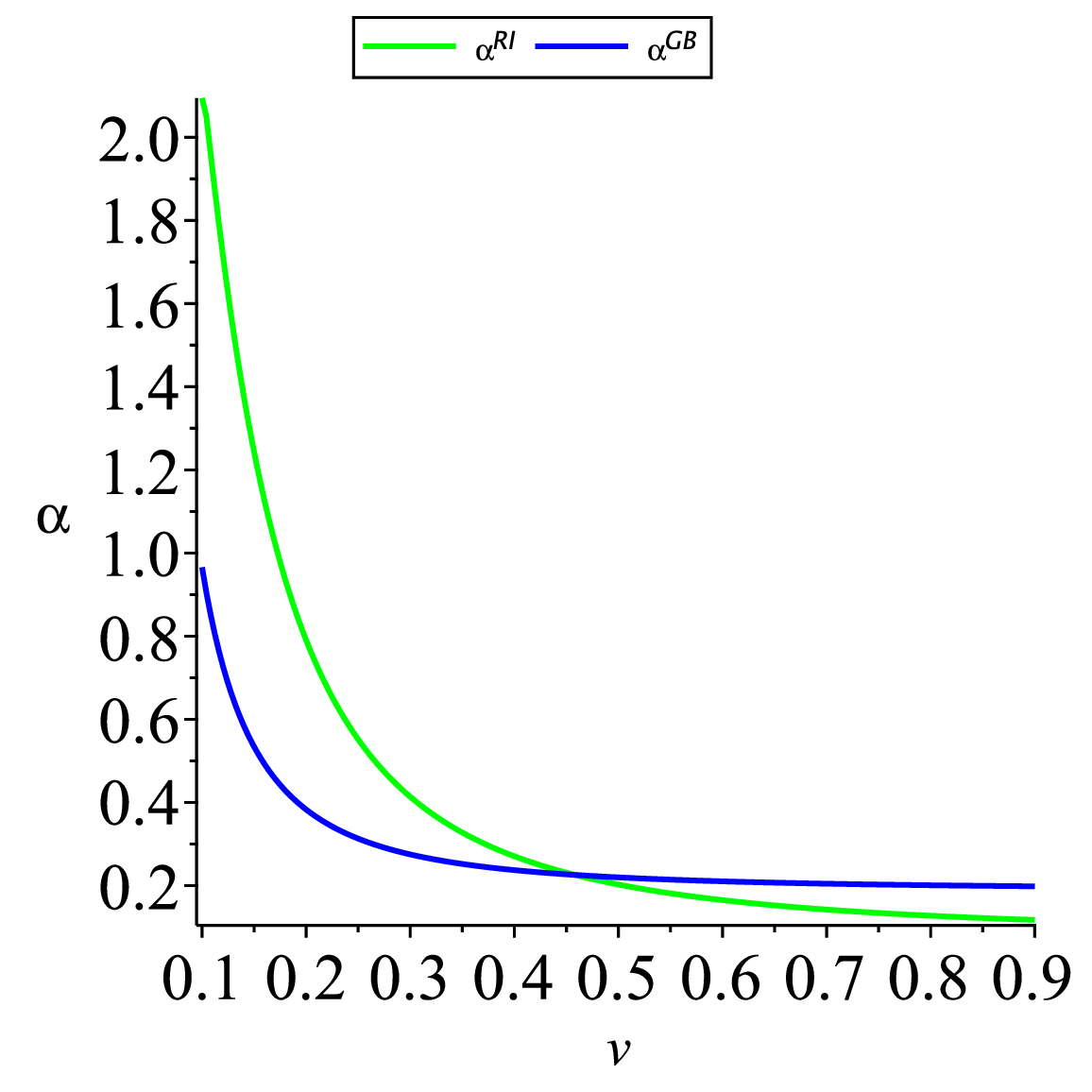}
\includegraphics[scale=0.3]{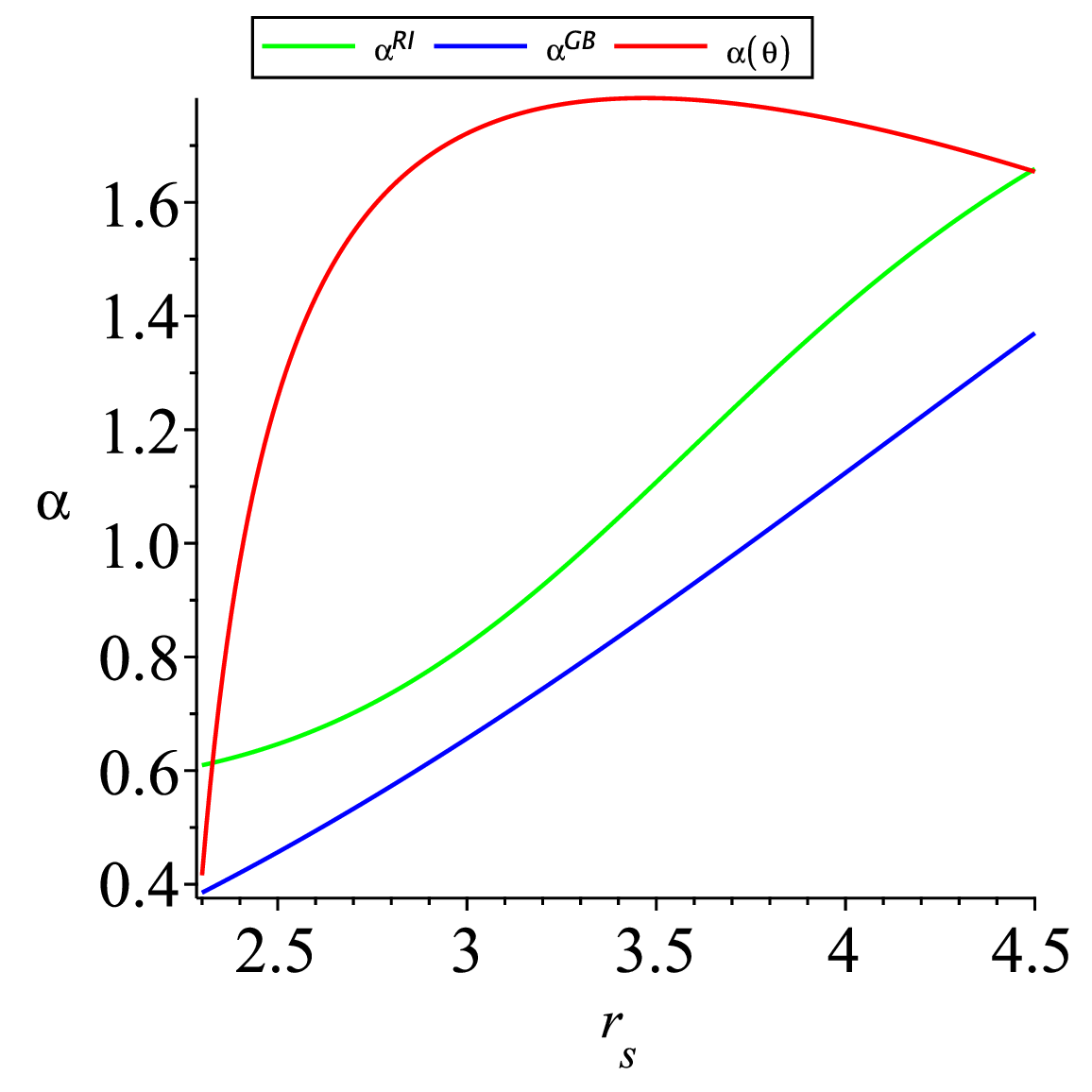}

\caption{The diagrams above provide a graphical difference of the deflection angle $\alpha$ for the Rindler-Ishak method and the Gauss-Bonnet theorem, shown as functions of the parameters $b$ (left part),
$v$ (middle part), and $r_s$ (right part) for different values of $\phi$. Additionally, for the strong field approximation, the deflection angle
$\alpha(\theta)$ is shown as $r_s$ dependent (right part),when  $ v \rightarrow 1$ . Here we assume that $Q=1.5$.}
\end{figure}

\begin{figure}[h]
\includegraphics[scale=0.3]{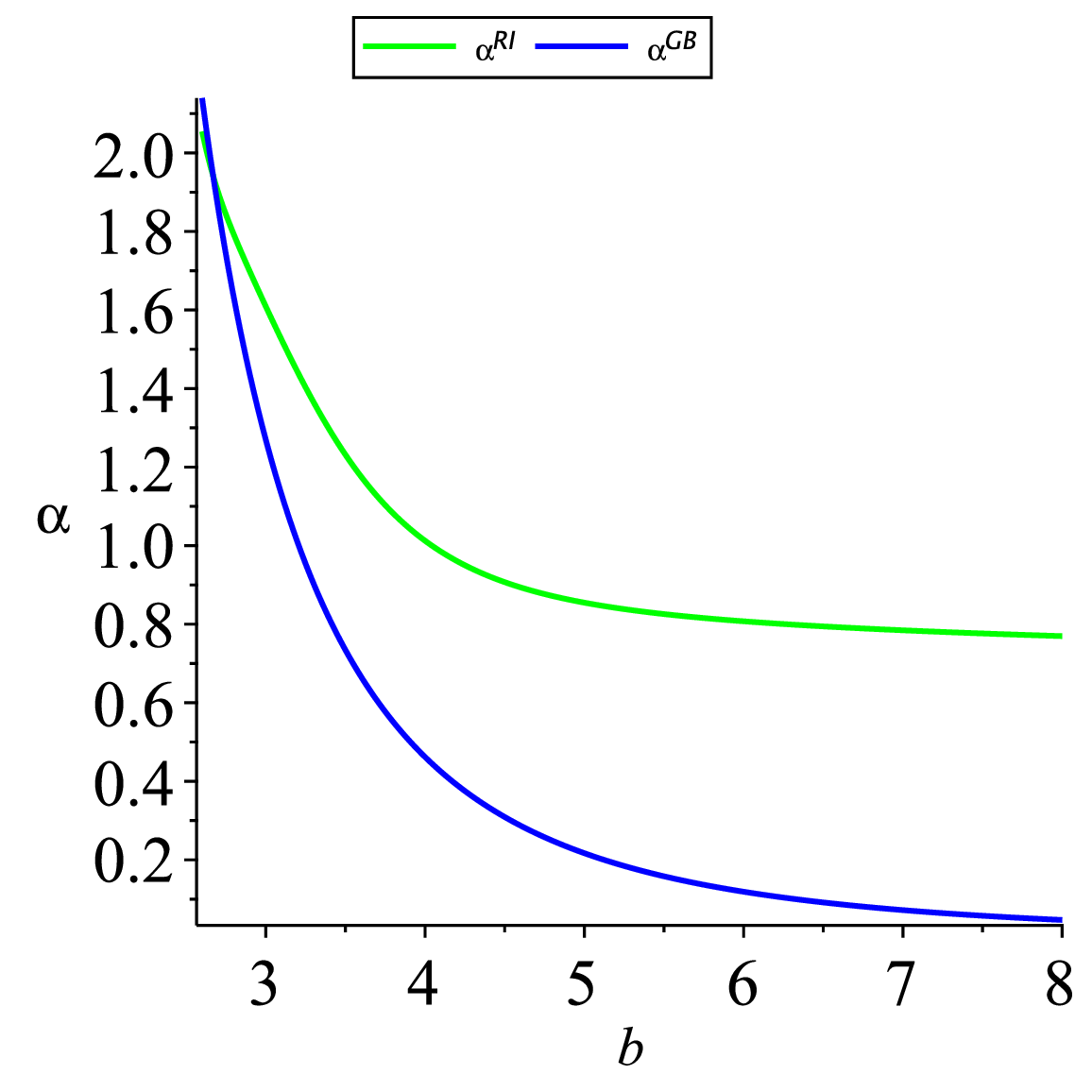}
\includegraphics[scale=0.3]{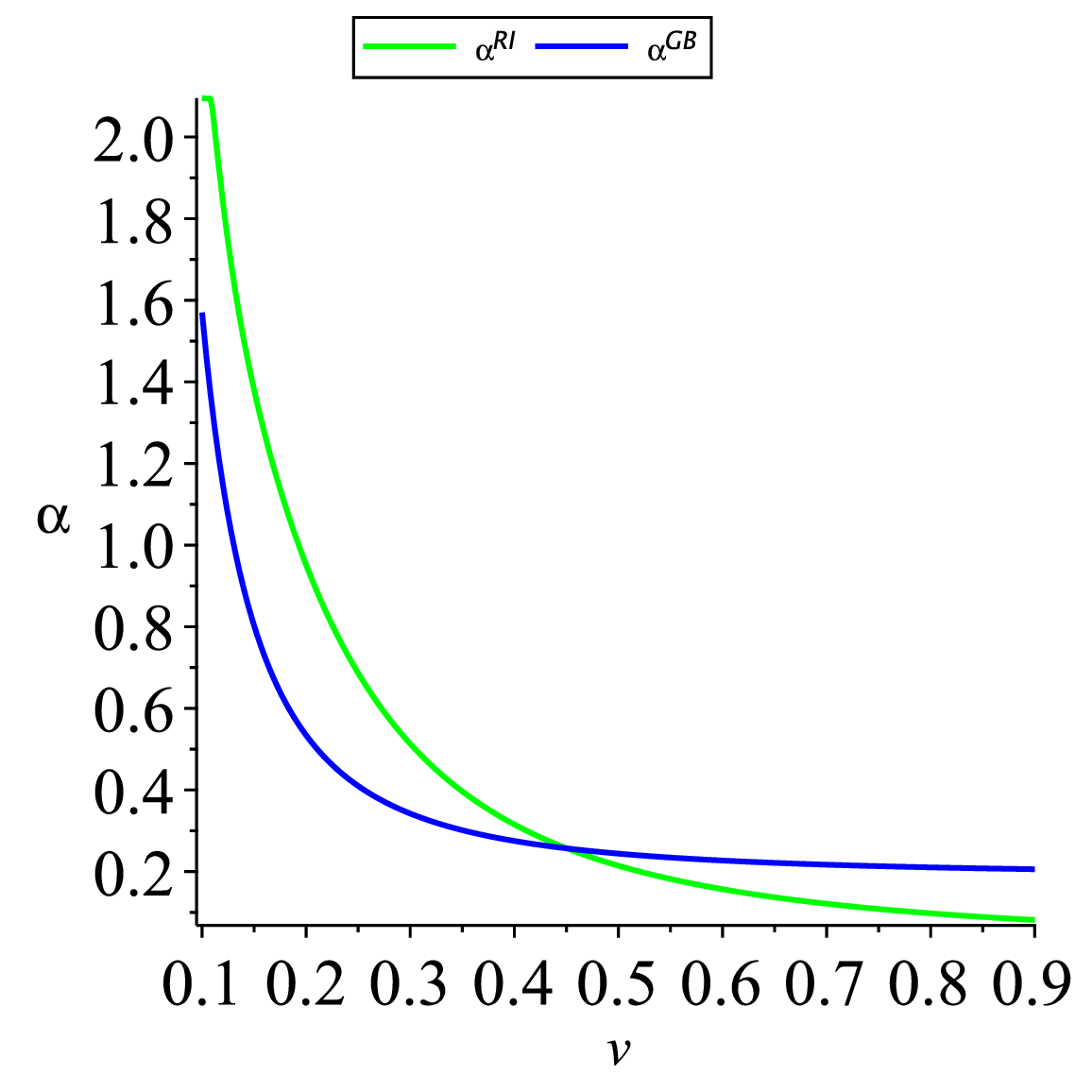}
\includegraphics[scale=0.3]{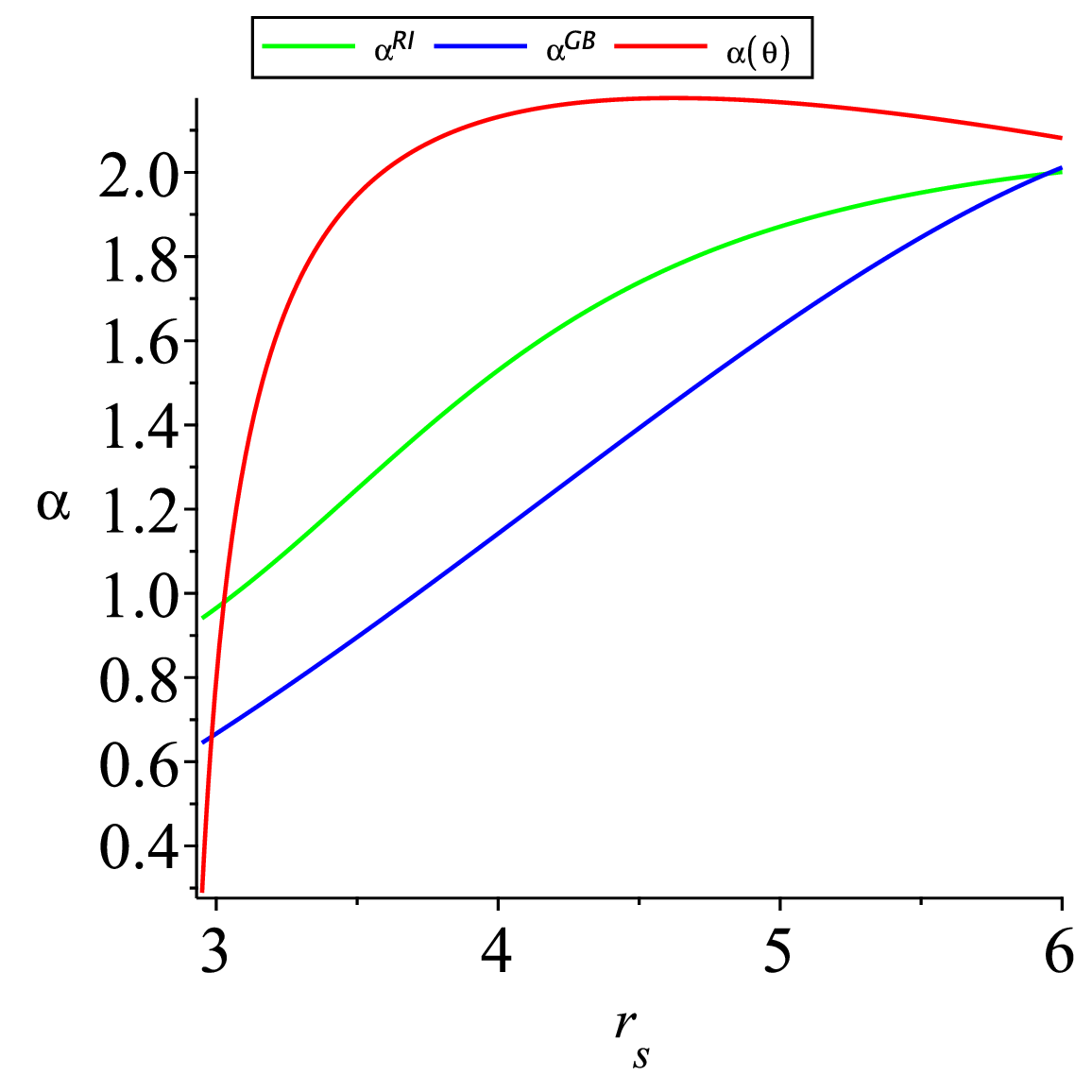}

\caption{The diagrams above provide a graphical difference of the deflection angle $\alpha$ for the Rindler-Ishak method and the Gauss-Bonnet theorem, shown as functions of the parameters $b$ (left part),
$v$ (middle part), and $r_s$ (right part) for different values of $\phi$. Additionally, for the strong field approximation, the deflection angle
$\alpha(\theta)$ is shown as $r_s$ dependent (right part), when  $ v \rightarrow 1$. Here we assume that $Q=2$.}
\end{figure}

 {
Recent observable data has hinted at the existence of black holes at the centers of spiral galaxies, prompting numerous researchers to propose models modifying Einstein's gravity to account for wormholes in different regions of galaxies. While the scientific community continues to seek observable evidence supporting the existence of wormholes, our study offers a fresh approach by incorporating Yoshiaki Sofue's exponential dark matter density profile. This new density profile leads to a novel theory of charged galactic wormholes, and our solutions fulfill all the criteria required for wormhole formation, suggesting that such structures could indeed exist. The charged wormhole metric we developed represents a significant advancement in understanding galactic dynamics, integrating charge effects within a realistic dark matter distribution and revealing unique interactions between dark matter, electromagnetic fields, and spacetime curvature. This model may predict distinctive gravitational phenomena, such as asymmetric Einstein rings, chromatic lensing, and Shapiro time delays \cite{x1}, offering new avenues for observational exploration. The convergence of results from the Gauss-Bonnet and Rindler-Ishak methods for gravitational lensing further validates the model's reliability, which will be our new project. These theoretical predictions can be tested with advanced observational instruments such as the James Webb Space Telescope, ALMA, or gravitational wave detectors, bridging the gap between theoretical constructs and empirical verification. This research opens the door to a deeper understanding of dark matter, electromagnetic fields, and spacetime geometry, providing a complementary framework for studying exotic astrophysical objects like charged wormholes, and offering insights that traditional mass-based models cannot fully capture.}

\pagebreak

\section{Acknowledgments:}
We are thankful to the referees for their constructive suggestions which help a substantial improvement of the manuscript.
FR and MKH would like to thank the authorities of the Inter-University Centre for Astronomy and Astrophysics, Pune, India
for providing the research facilities.     FR also gratefully acknowledges for financial support by DST-SERB, Govt. of India.

\end{document}